\documentclass[preprints,article,accept,pdftex,moreauthors]{Definitions/mdpi} 
\firstpage{1} 
\pubvolume{1}
\issuenum{1}
\articlenumber{0}
\pubyear{2023}
\copyrightyear{2022}
\externaleditor{Academic Editor: {Anna Trubetskaya }}

\datereceived{} 
\daterevised{} % Only for the journal Acoustics
\dateaccepted{} 
\datepublished{} 
\hreflink{https://doi.org/} % If needed use \linebreak
\usepackage[labelformat=simple]{subcaption}

\DeclareCaptionLabelFormat{subcaptionlabel}{\normalfont(\textbf{#2}\normalfont)}
\usepackage{enumitem}
\usepackage{placeins}
\usepackage{bm}
\Title{Assessing Long-Term Medical Remanufacturing Emissions with Life Cycle Analysis}

\TitleCitation{\linebreak Assessing Long-Term Medical Remanufacturing Emissions with Life Cycle Analysis}
\Author{Julia A. Meister \orcidA{}, Jack Sharp, Yan Wang *\orcidD{} and  Khuong An Nguyen \orcidC{}}

\AuthorNames{Julia A. Meister, Jack Sharp, Yan Wang and Khuong An Nguyen}

\AuthorCitation{Meister, J.A.; Sharp, J.; Wang, Y.; Nguyen, K.A.}
\address[1]{University of Brighton, School of Architecture, Technology and Engineering, Brighton BN2 4AT, UK; j.meister@brighton.ac.uk (J.A.M.); j.sharp1@uni.brighton.ac.uk~(J.S.); k.a.nguyen@brighton.ac.uk (K.A.N.)}

\corres{Correspondence: y.wang5@brighton.ac.uk}

\abstract{{The unsustainable take-make-dispose linear economy prevalent in healthcare contributes 4.4\% to global Greenhouse Gas emissions. A popular but not yet widely-embraced solution is to remanufacture common single-use medical devices like electrophysiology catheters, significantly extending their lifetimes by enabling a circular life cycle. To support the adoption of catheter remanufacturing, we propose a comprehensive emission framework and carry out a holistic evaluation of virgin manufactured and remanufactured carbon emissions with Life Cycle Analysis (LCA). We followed ISO modelling standards and NHS reporting guidelines to ensure industry relevance. We conclude that remanufacturing may lead to a reduction of up to 60\% per turn ($-$1.92 kg CO$_2$eq, burden-free) and 57\% per life ($-$1.87 kg CO$_2$eq, burdened). Our extensive sensitivity analysis and industry-informed buy-back scheme simulation revealed long-term emission reductions of up to 48\% per remanufactured catheter life ($-$1.73 kg CO$_2$eq). Our comprehensive results encourage the adoption of electrophysiology catheter remanufacturing, and highlight the importance of estimating long-term emissions in addition to traditional emission metrics.}}


\begin{document}

\section{Introduction}\label{sec:introduction}

    Global healthcare services contribute an enormous 4.4\% to the world's Greenhouse Gas emissions, enough to place it fifth in the country carbon footprint ranking just after China, USA, India, and Russia~\cite{karliner2019health}. A large portion of these emissions comes from waste processes. For example, 590,000 tonnes of healthcare waste are produced annually in England alone, creating significant environmental impacts and a financial cost of over \textsterling700 m~\cite{zils2022accelerating}. In 2020, the NHS became the world's first health system committed to a legally binding net zero emissions target~\cite{gov2022health}, recognising and solidifying the importance of reducing carbon emissions in healthcare.
    
    The NHS Net Zero plan and similar global initiatives acknowledge that the current take-make-dispose linear economy marked with single-use devices (SUDs) is a significant contributor to climate impact~\cite{england2020delivering}. Fortunately, there are promising opportunities for improvement. A circular economy approach to retain value through reuse, remanufacturing, and recycling reduces premature device disposal and may lead to large emission and financial savings~\cite{van2021new,cole2018reverse,guzzo2020circular}. However, remanufacturing comes with considerable challenges, as regulations understandably require that SUD medical devices are `reset' before clearance for reuse (i.e., returned to a state substantially equivalent to a new device)~\cite{mhra2016remanufacture,mhra2021reprocess}.

    Because of the stringent healthcare requirements, the medical remanufacturing process requires a series of resource-heavy steps (e.g., cleaning, sterilisation, testing)~\cite{schulte2021combining}. Therefore, careful consideration of the environmental impacts is required to ensure that it would indeed reduce emissions compared to producing a new device. To support decisions regarding linear versus circular procurement strategies, Life Cycle Analysis (LCA) is often employed to systematically assess the environmental impacts~\cite{liu2022benefit}. To ensure reliable results, comprehensive industry standards have been developed in recent years (e.g., ISO standards~\cite{iso200614040} and NHS reporting guidelines~\cite{penny2012greenhouse,standard2013ghg}).
    
    For remanufacturing to become a widely accepted practice, a collaboration between industry, legislation, and healthcare institutions is necessary. Previous studies have shown that successful remanufacturing is driven by engagement and support from all participating actors, who are largely motivated by triple bottom line benefits: profit, people, and planet~\cite{vadoudi2022comparing,jensen2019creating}. With that in mind, in this article, we focus on evaluating the long-term emission savings of remanufactured electrophysiology catheters. They are a promising candidate for remanufacturing on a large scale because they have the potential for significant financial savings (up to \textsterling1.7 m annually in the UK~\cite{leung2019remanufactured}), physician motivation is high (62\% in 42 EU-based healthcare centres~\cite{boussuge2022current}), and remanufacturing is promising (40\% of ablation procedure emissions come from catheters~\cite{ditac2022carbon}).

   In this article, we explore three research questions related to a circular remanufacturing approach for electrophysiology catheters:
    \begin{enumerate}[label=(\roman*)]
        \item {Can we validate the EP catheter remanufacturing results achieved by the Fraunhofer case study~\cite{schulte2021combining}? (Section~\ref{sec:burdenfreeResults})}
        \item {To what extent do key life cycle parameters affect the overall emission results? (Section~\ref{sec:sensitivityAnalysis})}
        \item {How can we incorporate a realistic, industry-informed circular use of catheters to evaluate long-term emission savings? (Section~\ref{sec:buyBack})}
    \end{enumerate}

    \subsection*{Contribution}
       
        The novelty of our work is the comprehensive and contextualised analysis of Life Cycle Analysis (LCA) emissions of medical single-use electrophysiology catheters, including the proposal of a long-term emission metric. Following on from the research questions laid out above, we identify three main objectives that this article addresses:
    
        \begin{enumerate}[label=(\roman*)]
            \item \emph{To validate a previous case study on the environmental emissions of electrophysiology catheter remanufacturing}~\cite{schulte2021combining} by the prestigious Fraunhofer Institute. \\
            {We have developed a sophisticated, industry-informed Life Cycle Analysis model with the open source openLCA software~\cite{di2019openlca} for both virgin manufactured and remanufactured catheters (Section~\ref{sec:lci}).}
            
            \item \emph{To perform a sensitivity analysis of key life cycle parameters, showcasing the magnitude of impact that model uncertainty may have on the total emission results}. \\
            After carrying out a holistic evaluation of virgin and remanufactured catheter emissions, {we analysed the impact of three key circular economy life parameters: the remanufacturing location, the catheter rejection rate, and the number of remanufacturing turns} (Section~\ref{sec:ghgResults}).
            
            \item \emph{To propose a novel framework to assess the realistic, long-term emissions of adopting a circular economy approach with remanufactured medical devices.} \\
            Our proposed framework models an industry-informed buy-back scheme to evaluate long-term remanufacturing emissions (Section~\ref{sec:buyBack}). We interpret the results in a healthcare context, taking industry stances and existing literature into account (Sections~\ref{sec:discussion} and~\ref{sec:relatedWork}).
        \end{enumerate}

\section{{Methodology}}\label{sec:methodology}
   
\begin{figure}[H]
        \centering
        \includegraphics[width=1\textwidth]{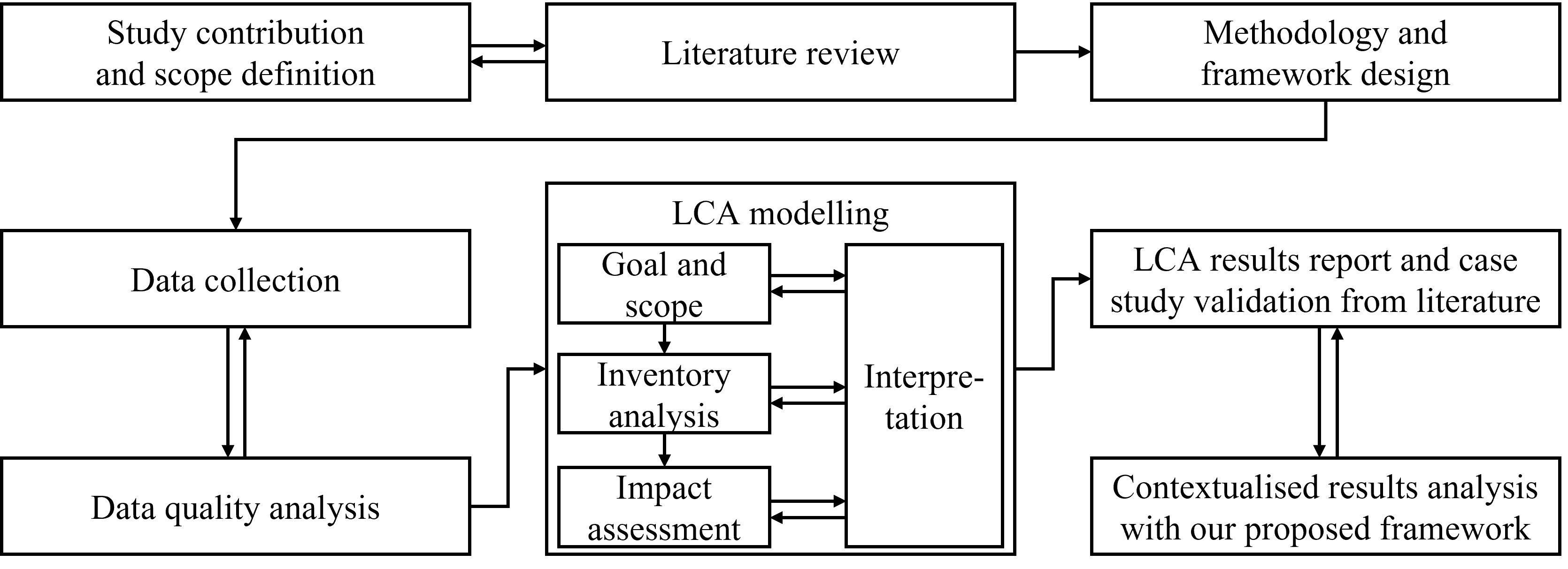}
        \caption{\emph{Methodology flowchart.}
        Our proposed framework extends existing industry standards to ensure that it may be seamlessly adopted by industry practitioners. Therefore, ISO regulations 14040~\cite{iso200614040} and NHS emission reporting standards~\cite{penny2012greenhouse,standard2013ghg} have significantly influenced our methodology, LCA models, and result reports.}\label{fig:flowchart}
    \end{figure}

    {As discussed in detail in Section~\ref{sec:introduction}, the purpose of this study is to develop a novel emission framework to assess the realistic, long-term emissions of a circular medical device economy. To evaluate our approach, we test the proposed framework on a use case of remanufactured single-use electrophysiology catheters and compare our results against a state-of-the-art emissions analysis in the literature. To estimate the emissions of a single medical device, we use Life Cycle Analysis (LCA) modelling, a systematic tool to analyse a product's environmental impacts over its entire life cycle~\cite{kloverpris2018establishing}. This makes our framework device-agnostic, and the comprehensive analysis of single-use catheter remanufacturing may be extended to other medical devices in the future. A flowchart of our methodology is presented in Figure~\ref{fig:flowchart}.}

    {To ensure that our comprehensive emission analysis framework maintains a high standard and may be seamlessly adopted by the industry for climate impact analysis, we incorporate widely-accepted industry guidelines. The most impactful LCA standards are ISO 14040~\cite{iso200614040} and 
    NHS regulations~\cite{penny2012greenhouse,standard2013ghg}. More details are given in Sections~\ref{sec:lcaMethod}.}

    {The following sections discuss the methodology of data collection, LCA modelling, and emission metrics in more detail.}

    \subsection{{Life Cycle Analysis Methodology}}\label{sec:lcaMethod}
       
        {In this article, we carry out a comparative analysis of a virgin manufactured and a remanufactured electrophysiology catheter's climate impact. We use Life Cycle Analysis (LCA) modelling to assess a single product's emissions over the course of its life. LCA systematically quantifies the climate impact of each material and waste input and output. Subsequently, the LCA results are used as the input to our proposed emission framework, which contextualises the emission savings by calculating the realistic, long-term climate impacts of a circular catheter economy. In this section, we present a robust methodology for a comparative LCA. All models and impact calculations were carried out in \emph{openLCA} version 1.10.3, an open-source LCA software~\cite{di2019openlca}.}

        {Our LCA modelling approach was heavily influenced by industry regulations. In particular, we incorporated ISO 14040~\cite{iso200614040} guidelines. Additionally, NHS England has published numerous sophisticated LCA reporting and modelling standards~\cite{penny2012greenhouse,standard2013ghg} which were taken into account. These standards are highly significant because they encode the unified principles of a thorough LCA impact analysis, define relevant terms, and describe best practices. The benefit of incorporating these guidelines is that they are defined on an international and national level, respectively, and therefore ensure that the LCA modelling is of high quality and comparable between studies.}
        
        \clearpage
        {ISO  prescribes four essential and interconnected steps for LCAs (also visualised in Figure~\ref{fig:flowchart}): Goal and Scope definition (Section~\ref{sec:methodology}), Inventory analysis (Section~\ref{sec:lci}), Impact assessment (Section~\ref{sec:ghgResults}), and Results interpretation (Sections~\ref{sec:ghgResults} and \ref{sec:discussion}). In contrast, the NHS guidelines are particularly relevant for the emission analysis of medical devices. They give detailed rubrics for the quantitative life cycle materials that should be included in the LCA (see Table~\ref{tab:includedFlows}), and the qualitative assessment of LCA data quality (see Section~\ref{sec:dataQuality}).}
        
        {Following the guidelines discussed above, we take the \emph{cradle-to-grave} modelling approach to measure the impact of a catheter over its entire life~\cite{botejara2022comparative}. For a virgin, newly manufactured catheter, this means from the production of the raw resources (cradle) to incineration after one use (grave). For a remanufactured catheter, we measure emissions from the pickup of a used catheter, e.g., at a hospital, to when it has been reused after a successful remanufacturing cycle (grave). The \emph{system boundary} describes which attributable and non-attributable processes are included in the emission calculations. We follow the quantitative material input guidelines for medical devices as described in the NHS GHG accounting guidance~\cite{penny2012greenhouse}. The list of included and excluded material flows is given in Table~\ref{tab:includedFlows}. Details of the individual life stages and material inputs are described in Section~\ref{sec:lci}.}

        \begin{table}[H]
            \centering
            \caption{\emph{Quantitative rubric of the relevant material inputs and outputs for medical device LCA models.
            Taken from the NHS guideline for accounting and reporting medical device life cycles\cite{standard2013ghg}}.}\label{tab:includedFlows}
                \begin{tabularx}{\textwidth}{XXX}
                    \toprule
                    \multicolumn{2}{c}{\textbf{{Included Processes}}} & \multicolumn{1}{c}{\textbf{{Excluded Processes}}} \\
                   \midrule
                    \textbf{{Attributable Processes}} & \textbf{{Non-Attributable   Processes}} & \textbf{{Attributable \& Non-Attrib.}} \\
                    \midrule
                    {Production,   processing and transport of raw materials.} & {Production and distribution of   energy/water/chemicals for sterilisation of reusable devices.} & {Transport of staff involved in   delivery, maintenance, refurbishment and repair.} \\\midrule
                    {Manufacture,   sterilisation, packing, storage and distribution of the device.} & {Production and distribution of   consumables required for the operation of a medical device.} & {Transport of patients to receive   treatment.} \\\midrule
                    {Production   and distribution of energy, water and materials consumed by the medical   device during operation.} & {Sterilisation and cleaning   chemicals.} & {General hospital/clinic/home   infrastructure to support use of the medical device.} \\\midrule
                    {Production   of spare parts/materials, and energy required for refurbishment and repair of   devices.} & {Refrigerant leakage associated   with product manufacturing}. & {Software to run medical devices.} \\\midrule
                    {Waste   management activities at end-of-life, including transportation.} &  & {Ancillary products and   equipment, e.g., protective clothing.} \\
                    \bottomrule
                \end{tabularx}
        \end{table}
        
        For a meaningful comparison between one virgin and one remanufactured catheter's life, we choose the \emph{cut-off system model}. Consequently, emissions from waste treatment after the first life are attributed to the producer~\cite{baustert2022integration}. In other words, the used virgin catheters are burden-free for the remanufacturing process.
        
        Finally, we forgo \emph{normalisation} and \emph{weighting} for the final LCA emission calculations. These techniques may be used to aggregate impact effects and evaluate the results relative to reference information~\cite{sanchez2021life}, while there are benefits to such techniques, they are marked as optional in LCA ISO guidance~\cite{borglin2021life} due to widespread criticism of their usefulness. We skip these steps to avoid inadvertently introducing bias through the choice of reference information and to maintain interpretability~\cite{pizzol2017normalisation}.
        
  \clearpage  
    \subsection{{System and Unit-Level Assumptions}}\label{sec:assumptions}
       
        {The more detailed, expansive, and specific the underlying process data is, the more accurate and realistic a Life Cycle Analysis (LCA) model will be~\cite{salemdeeb2021pragmatic}. However, in the interest of time and cost-effectiveness, most LCA regulations acknowledge a necessary balance between model exactness and simplicity~\cite{iso200614040,iso200614044,penny2012greenhouse,standard2013ghg}, while defining universally effective simplification methods is still an open problem, \cite{gradin2021common} suggests that meaningful LCA analysis should include a detailed description of any assumptions built into the model. To increase the interpretability of our results, the following list summarises the system-level assumptions we made.}
        \begin{enumerate}[label=(\roman*)]
            \item Catheter producers and remanufacturers follow a Just-in-Time (JIT) inventory management system. Therefore, catheters do not need to be stored, and there are no emissions tied to the running and maintenance of a warehouse.
            \item {Since OEMs produce many virgin device models, we assume that our generalised virgin device is representative of them. This position is supported by our industry~collaborators.}
            \item {\textls[-15]{We assume that the modelled virgin single-use device is eligible for remanufacturing.}}
            \item We assume that all catheter production steps (e.g., production, assembly, packaging) are performed in the same manufacturing facility and, therefore, transportation of individual components is not required. Equivalently, all remanufacturing steps (e.g., disassembly, cleaning, sterilisation) are performed in the same remanufacturing facility.
            \item Due to concern over the potential contamination of used medical equipment, we assume that virgin and remanufactured end-of-life devices are disposed of by incineration (e.g., rather than recycled).
            \item We assume that used catheters arriving at a remanufacturing facility are fully functional and do not require replacement components.
            \item We assume that chips embedded in catheters are not locked by the OEM after first use and therefore do not need to be unlocked during remanufacturing.
        \end{enumerate}
        
        {We also make the following unit-level assumptions. Concrete details are given in the description of relevant life cycle stages in Section~\ref{sec:lci} for both virgin and remanufactured single-use devices.}
        \begin{enumerate}[label=(\roman*)]
            \item {In cases where proprietary component plastics are not available in the Ecoinvent dataset, we assume that replacing these with plastics that have similar production characteristics (most importantly, melting point) will have a similar overall climate~impact.}
            \item {We assume that some life cycle processes are carried out manually, and therefore do not include material inputs or outputs for those stages.}
            \item {We assume that the virgin and remanufactured catheters' packaging and use stages are the same.}
            \item {We assume that ready-to-use catheters (virgin and remanufactured) are packaged~manually.}
            \item {We assume that ready-to-use catheters do not require regular maintenance because they are single-use devices.}
            \item {We assume that the average use duration of a catheter in surgery is 1 h, supported by primary data from AMDR.}
            \item {We assume that the waste-disposal facility is in the vicinity of the user, and therefore exclude transportation to the incineration site.}
        \end{enumerate}
        
        Our system and unit-level assumptions may lead to slightly conservative emissions. We take this into account with a sensitivity analysis in Section~\ref{sec:sensitivityAnalysis} and an in-depth discussion of our results in Section~\ref{sec:discussion}.

     \subsection{Data Sources and Quality Assessment}\label{sec:dataQuality}
       
        Life Cycle Analysis (LCA) models are made up of foreground and background data. Foreground data is specific to the product system that is modelled, while background data gives the context of the wider economic and technological systems~\cite{steubing2021making}. To achieve a realistic representation of the product systems, we collected primary data where possible and filled in gaps from secondary sources. The following data sources were consulted:
            \begin{enumerate}[label=(\roman*)]
                \item {\emph{Primary foreground sources.}
                NHS England and USA-based medical device remanufacturers AMDR and Innovative Health provided primary process activity data from 2021. \\
                This data focused on life cycle statistics influenced by actors across the single-use catheter's value chain. In particular, parameters such as the remanufacturing collection, rejection, and success rates. Additionally, our collaborators validated the secondary material and process data for the catheter components and life cycle processes.
                \item {\emph{Secondary foreground sources.} A 2021 case study~\cite{schulte2021combining} on the life cycle impacts of electrophysiology catheter remanufacturing conducted by the prestigious Fraunhofer Institute for Environmental, Safety and Energy Technology was the source of secondary process activity. \\
                The study provides a detailed bill of materials of the single-use catheter, provided by the Germany-based medical remanufacturer Vanguard AG. Some simplifications were made, including neglecting a filler of polyurethane in the catheter curvature and loop because details about the concentration (and therefore mass) were unknown. However, the authors suggest that this did not significantly influence the total emissions, as the curvature and loop represent only 0.7\% of the catheter's total weight. Similarly, when proprietary component materials were not available in the background dataset, the study replaced them with possible preliminary products (e.g., for the catheter shaft). The article reports that the recorded remanufacturing processes are in full compliance with the specifications of the original product and with the EU Medical Device Directive 93/42/EEC.}
                \item {\emph{Background data sources}. The established industry-standard Ecoinvent v3.8 dataset~\cite{wernet2016ecoinvent}} provided background data and emission factors for the product processes. \\
                Data points were collected between 2006 and 2011. Ecoinvent process details are given for each material flow data point when discussing the virgin and remanufactured catheter Life Cycle Inventory (LCI) in Section~\ref{sec:lci}.}
            \end{enumerate}
            
        Since the quality of a Life Cycle Impact Assessment (LCIA) is directly correlated with the quality of the product system’s underlying data~\cite{edelen2018creation}, we report the data quality for each input and output flow in Section~\ref{sec:lci}. Following the NHS guideline for life cycle reporting~\cite{penny2012greenhouse} we used an adjusted qualitative data quality appraisal technique (Table~\ref{tab:dataQuality}).
       
        {Furthermore, data precision is key in LCIAs, as impacts may be highly sensitive to even minor changes in a range of factors~\cite{schulte2021combining}. Where details about process region, production details, and producers were reported, they were included in the models. Where specific information was missing or inconclusive, we used relevant alternatives to improve data quality following best practices. Details are provided as part of the LCI in Section~\ref{sec:lci} for all affected material flows and life cycle stages. Improvement methods included:}
        \begin{enumerate}[label=(\roman*)]
            \item When data ranges were given, we calculated average values.
            \item When material producer information was not given, we selected producers from regions with similar production conditions.
            \item Proprietary plastics that were not represented in the Ecoinvent~\cite{wernet2016ecoinvent} dataset were replaced with alternatives. When selecting the best alternative, the focus was on matching characteristics relevant to plastic production, as plastic processing has the highest environmental impacts throughout a plastic’s life~\cite{mannheim2021life}.
        \end{enumerate}
        
        \begin{table}[H]
            \caption{\emph{Qualitative data quality appraisal rubric.} Adjusted from the NHS guideline for accounting and reporting medical device life cycles~\cite{standard2013ghg}.}\label{tab:dataQuality}
            \begin{adjustwidth}{-\extralength}{0cm}
                \begin{tabularx}{\fulllength}{lXllXXX}
                    \toprule
                    \textbf{Score}
& \textbf{Technology} & \textbf{Age} & \textbf{Geography} & \textbf{Completeness} & \textbf{Reliability} & \textbf{\emph{Note}} \\
                    \midrule
                    \textbf{Very good} & Consistent generation. & <3 years & Same area. & All relevant sites. & Verified data based on measurements. & \emph{Primary data from AMDR, Innovative Health, and NHS England.} \\\midrule
                    \textbf{Good} & Similar generation. & 3--6 years & Similar area. & More than 50\% of sites for an adequate time. & Non-verified data based on measurements. & \emph{Data taken directly from the Fraunhofer case study.} \\\midrule
                    \textbf{Fair} & Different generation. & 6--10 years & Different areas. & \mbox{Less than 50\% of} \mbox{sites for an adequate} time. & Non-verified data partly based on assumptions or a qualified estimate. & \emph{Data adapted from the Fraunhofer case study (e.g., due to data availability).} \\\midrule
                    \textbf{Poor} & Unknown generation. & >10 years & Unknown area. & \mbox{Less than 50\% of} \mbox{sites for a shorter} time. & Non-qualified estimate. & \emph{Not applicable.} \\
                    \bottomrule
                \end{tabularx}
            \end{adjustwidth}
        \end{table}
        
    \subsection{{Sensitivity Analysis Methods}}\label{sec:sensitivityParameters}
        {As discussed in Section~\ref{tab:dataQuality}, Life Cycle Analysis (LCA) results may be sensitive to even minor changes in the input data. Consequently, sensitivity analysis is a valuable tool to assess uncertainty in the environmental impact results~\cite{mcalister2021lca}. In collaboration with NHS England and industry remanufacturers AMDR and Innovative Health, three high-impact remanufacturing parameters were identified for analysis:}
        \begin{enumerate}[label=(\roman*)]
            \item {The \emph{remanufacturing location} $L$ dictates the transportation mode and route of the used and remanufactured catheters between the user and remanufacturing facilities (studied $L \in \{$DE, UK, USA$\}$).}
            \item {The \emph{rejection rate} $R$ describes the percentage of used catheters discarded during remanufacturing due to their sub-par quality (studied $0\% \leq R \leq 70\%$).}
            \item {The \emph{number of turns} $N$ identifies the number of times a catheter may be used overall. For example, a catheter with $N=3$ turns is used once as a virgin catheter and twice as a remanufactured catheter before reaching the end of its life (studied $1 \leq N \leq 5$).}
        \end{enumerate}

        {To assess the importance of the above life cycle parameters, we carried out a comprehensive \emph{univariate sensitivity analysis} in Sections~\ref{sec:sensitivtyRemanLocation}--\ref{sec:sensitivityNrTurns}, varying one of the three parameters at a time. After analysing the individual magnitude of LCA result changes, we evaluated the combined effect of the parameter changes in three representative remanufacturing scenarios. The \emph{multivariate sensitivity analysis} results are presented in Section~\ref{sec:sensitivityScenarios}.}
        \begin{enumerate}[label=(\roman*)]
            \item {\emph{Good scenario}: Ideal scenario ($L=$UK, $R=0$\%, $N=5$).}
            \item {\emph{Average scenario}: Approximately current ($L=$DE, $R=15$\%, $N=4$).}
            \item {\emph{Bad scenario}: Approximately following~\cite{schulte2021combining} ($L=$USA, $R=50$\%, $N=3$).}
        \end{enumerate}

    \subsection{Emission Metrics}\label{sec:metrics}    
        We chose \emph{climate impact} as the primary indicator of a catheter's effects on the environment. Furthermore, known as Greenhouse Gas (GHG) emissions or carbon footprint, climate impact is central to many widely-circulated technical reports and regulations (e.g., the legally binding NHS net zero commitment~\cite{england2020delivering}, EU Commissions Climate Target Plan~\cite{eu2020carbon}, and the UN's Paris Agreement~\cite{agreement2015paris}, to name a few). Consequently, the term is familiar to a global audience across the public, industry, and political spectrum.

        To quantify a catheter's climate impact, we followed the Environmental Footprint (EF) 3.0 methodology, a European Commission initiative to standardise the calculation of emissions during a product's life cycle~\cite{eu2013ef}. The unit is kg CO$_2$eq (spoken ``carbon dioxide equivalent''), which allows different GHGs to be reported as one value by multiplying their weight with their Global Warming Potential (GWP)~\cite{brander2012greenhouse}. {For this article, the functional unit of analysis is defined as the production and potential disposal of one catheter (virgin or remanufactured) after its use in a 1-hour procedure.}
        
        {An inherent challenge with circular life cycles such as catheter remanufacturing is how to attribute the total emissions across individual turns. To address this, we follow the NHS GHG reporting guidelines~\cite{penny2012greenhouse,standard2013ghg} on how to calculate per life and per turn emissions. We report the following values in Section~\ref{sec:ghgResults}:}
        \begin{enumerate}[label=(\roman*)]
            \item {Burden-free per turn emissions $E^f_v$ for virgin manufactured and $E^f_r$ for remanufactured catheters;}
            \item {Burdened per turn catheter emissions $E^b_{perTurn}$;}
            \item {Furthermore, burdened per life catheter emissions $E^b_{perLife}$.}
        \end{enumerate}
        
        {Equations~\eqref{eq:virginFree} and~\eqref{eq:remanFree} define the total emissions of one virgin catheter $E^f_v$ and one remanufactured catheter $E^f_r$ as the sum of all input emissions $e_i$ and output emissions $e_o$. The superscript $f$ indicates that the emission results are \emph{burden-free}; No previous lives are accounted for. For example, the remanufactured catheter's emissions $E^f_r$ do not include emissions from its first virgin life.
            \begin{align}
                E^f_v &= \sum e_i + \sum e_o \label{eq:virginFree} \\
                E^f_r &= \sum e_i + \sum e_o \label{eq:remanFree}
            \end{align}}
            
        {In contrast, results marked with the superscript $b$ include the emissions from previous lives, i.e., the emissions are \emph{burdened}. These values are dependent on the \emph{number of turns} $N$, which describes the total number of uses before a catheter reaches the end of its life and is discarded. For example, a catheter with $N=3$ turns is used exactly three times: once as a virgin catheter and twice as a remanufactured catheter. The formula for a catheter's emissions across its whole life $E^b_{perLife}$ is given in Equation~\eqref{eq:perLifeBurdened}.
        \begin{align}
            E^b_{perLife} = E^f_v + (N-1) \cdot E^f_r \label{eq:perLifeBurdened}
        \end{align}}
    
        {For an equivalent turn comparison between a virgin catheter's emissions $E^f_v$ and a burdened remanufactured catheter's emissions, we may calculate the relative per turn emissions $E^b_{perTurn}$ with Equation~\eqref{eq:turnBurdened}.
        \begin{align}
            E^b_{perTurn} = \frac{1}{N} E^b_{perLife} \label{eq:turnBurdened}
        \end{align}}

    \subsection{{Proposed Long-Term Emission Metrics}}\label{sec:longTermMetric}
        {Industrial medical device remanufacturing is often carried out as a third-party service, and a common approach is to treat it as a buy-back scheme. Evaluating the long-term emissions, therefore, requires estimating the necessary number of units (i.e., catheter uses) required for a period of time, and purchasing an initial injection of virgin devices to be remanufactured until either the quality fails, or the maximum number of turns is reached.}
        
        {To simulate the emission savings of a long-term buy-back scheme compared to purchasing only new virgin catheters, we must first calculate the necessary number of new catheters to prime the system.  Given details of the remanufacturing process, we can solve for the initial number of virgin catheters $C$ needed to achieve $U$ total catheter uses in a buy-back scheme:
        \begin{gather}\label{eq:catheterUses}
            U = \sum^{N}_{n=1} C \cdot (1-R)^{n-1}
        \end{gather}
        $C$ is dependent on the cleared number of turns $N$ (where $n=1$ is a catheter’s first virgin life) and the catheter rejection rate $R$. Assuming that all used catheters are successfully collected and returned to the remanufacturer, we may calculate the initial injection of new devices. Similarly to $E^b_{perLife}$ and $E^b_{perTurn}$ in Equations~\eqref{eq:perLifeBurdened} and~\eqref{eq:turnBurdened}, we may define the long-term total $E^l_{perScheme}$ and per turn $E^l_{perTurn}$ emissions of the buy-back scheme.
        \begin{gather}
            E^l_{perScheme} = C \cdot E_v^f + (U-C) \cdot E_r^f \label{eq:longTermLife} \\
            E^l_{perTurn} = \frac{1}{U} \cdot E^l_{perLife} \label{eq:longTermTurn}
        \end{gather}}

\section{Single-Use Catheter Life Cycles and Material Flows}\label{sec:lci}
   
    Before analysing the environmental impacts, we provide a detailed Life Cycle Inventory (LCI). This illustrates the virgin manufactured and remanufactured catheter life cycle stages with a comprehensive description of material inputs and outputs. The start and end life cycle stages (i.e., cradle and grave) are dictated by the \emph{functional unit}. A functional unit is key in LCAs, as it defines one unit of the product or service being evaluated~\cite{demarco2021functional}. In this paper, we determine it as the production, use (1-h long procedure), and disposal of one catheter. By choosing equivalent functional units for both the virgin and remanufactured scenarios, we may confidently and intuitively compare their emission results (even though catheter `production' includes different processes)~\cite{furberg2022practice}.
    
    {To decide which inputs and outputs are contained in the system boundary, we use the quantitative rubric published by NHS guidelines (see Section~\ref{sec:dataQuality}). Additionally, we outline the quantity, source, and data quality of each data point in this section, following the NHS guidance for life cycle reporting~\cite{standard2013ghg}.}
    
    {Since component materials and processes may vary significantly based on the catheter model, Original Equipment Manufacturer (OEM), and remanufacturer, we selected representative product systems for the virgin and remanufactured life cycles. The inputs and outputs were largely informed by a 2021 EP catheter case study~\cite{schulte2021combining} which was supplemented and updated with our primary data from NHS England and USA-based medical device remanufacturers AMDR and Innovative Health (see Section~\ref{tab:dataQuality}). The primary data additionally influenced our catheter use and remanufacturing locations, which were set in the UK and the USA, respectively. To account for the possible bias introduced to our emission values by the chosen representation, we include a sensitivity analysis of key process parameters in our results analysis (Section~\ref{sec:ghgResults}).}

    \subsection{Virgin Manufactured Catheter}\label{sec:virginLifeCycle}
\begin{figure}[H]
            \centering
            \includegraphics[width=0.99\textwidth]{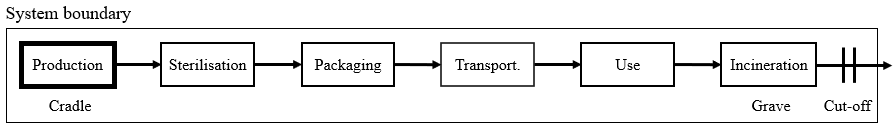}
            \caption{\emph{Virgin catheter product system.} The life stages are arranged in a linear life cycle from cradle-to-grave, which here is defined as production to incineration. We follow the cut-off system that is marked by the square system boundary.}\label{fig:virginLifeCycle}
        \end{figure}

        {A catheter’s linear life cycle was tracked from the production from raw materials in the USA (cradle) to its waste disposal after use in the UK (grave), as shown in Figure~\ref{fig:virginLifeCycle}. Because we selected the cut-off model (see Section~\ref{sec:lcaMethod}), any energy recovered after incineration was not credited to the catheter’s life cycle. Table~\ref{tab:virginMaterials} presents an overview of the process steps, and subsequent material flows per life stage.}
        
        \begin{table}[H]
            \caption{{\emph{Virgin life stages and materials}. The materials are based on~\cite{schulte2021combining} and supplemented with primary data from our collaborators, medical device remanufacturers AMDR and Innovative Health. The data quality was assessed with the rubric in Table~\ref{tab:dataQuality}, and the region marks the Ecoinvent production geography.}}\label{tab:virginMaterials}
            \begin{adjustwidth}{-\extralength}{0cm}
            \begin{tabularx}{\fulllength}{lrccccX}
                \toprule
                {\textbf{Stage/Material}} & {\textbf{Quantity}} & {\textbf{Source}} & {\textbf{Quality}} & {\textbf{Time Range}} & {\textbf{Region}} & {\textbf{Description}} \\
                \midrule
                {\textbf{Production}} &  &  &  &  &  & {\textbf{Component production from raw resources and assembly.}} \\

                {polyamide} & {$3.20 \times 10^3$ kg}
& {Adjusted~\cite{schulte2021combining}} & {Fair} & {2011--2021} & {GLO} & {Pre-product for the shaft.} \\
                {ethylene glycol} & {$1.25 \times 10^3$ kg} & {Adjusted~\cite{schulte2021combining}} & {Fair} & {2011--2021} & {RER} & {Pre-product for the shaft.} \\
                {polyethylene LD} & {$3.00 \times 10^4$ kg} & {\cite{schulte2021combining}} & {Good} & {2011--2021} & {GLO} & {Shaft stiffener.} \\
                {polysulfone} & {0.10810 kg} & {Adjusted~\cite{schulte2021combining}} & {Fair} & {2012--2021} & {GLO} & {Plug and handle.} \\
                {polyurethane} & {$0.80 \times 10^3$ kg} & {\cite{schulte2021combining}} & {Good} & {2018--2021} & {RER} & {Curvature and loop.} \\
                {electricity}	& {0.026 kWh} & {Adjusted Table~\ref{tab:remanMaterials}} & {Fair} & {2014-2021} & {US-WECC} & {Component assembly.} \\
                \midrule
                {\textbf{Sterilisation}} &  &  &  &  & & {\textbf{Gas sterilisation.}} \\
                {carbon dioxide} & {$2.82 \times 10^3$ kg} & {\cite{schulte2021combining}} & {Good} & {2011--2021} & {RER} & {Sterilisation gas ingredient.} \\
                {ethylene oxide} & {$1.80 \times 10^4$ kg} & {\cite{schulte2021combining}} & {Good} & {2018--2021} & {GLO} & {Sterilisation gas ingredient.} \\
                {electricity} & {0.360 kWh} & {\cite{schulte2021combining}} & {Good} & {2014--2021} & {US-WECC} & {Process electricity.} \\
                \midrule
                {\textbf{Packaging}} &  &  &  &  &  & {\textbf{Manual packaging.}} \\
                {polyethylene HD} & {$2.00 \times 10^2$ kg} & {\cite{schulte2021combining}} & {Good} & {2011--2021} & {GLO} & {Primary (sterile) packaging.} \\
                {carton box} & {$1.40 \times 10^1$ kg} & {Adjusted~\cite{schulte2021combining}} & {Fair} & {2011--2021} & {GLO} & {Secondary packaging.} \\
                \midrule
                {\textbf{Transport}} &  &  &  &  &  & {\textbf{Transport from manufacturer to user.}} \\
                \multirow{2}{*}{{container   ship}} & {0.11890 kg} & \multirow{2}{*}{\href{http://ports.com}{Ports.com}\textsuperscript{1}}
& \multirow{2}{*}{{Fair}} & \multirow{2}{*}{{2007--2021}} & \multirow{2}{*}{{GLO}} & \multirow{2}{*}{{USA-UK sea   route.}} \\
                 & {$\times$ 18,760 km}
&  &  &  &  &  \\
                \multirow{2}{*}{{lorry}} & {0.11890 kg} & \multirow{2}{*}{\href{https://www.google.co.uk/maps}{Google maps}\textsuperscript{1}} & \multirow{2}{*}{{Fair}} & \multirow{2}{*}{{2011--2021}} & \multirow{2}{*}{{RER}} & \multirow{2}{*}{{UK final distribution.}} \\
                 & {$\times$ 250 km} &  &  &  &  &  \\
                 \midrule
                {\textbf{Use}} &  &  &  &  &  & {\textbf{Materials consumed by device.}} \\
                {electricity} & {0.050 kWh} & {AMDR, IH, \cite{fda2007watts}} & {Very Good} & {2014--2021} & {GB} & {1 h of use, 50 watts.} \\
                \midrule
                {\textbf{Incineration}} &  &  &  &  &  & {\textbf{End-of-life waste disposal.}} \\
                {plastic incineration} & {0.11890 kg} & {\cite{schulte2021combining}} & {Good} & {2006--2021} & {CH} & {Municipal (100\%).} \\
                \bottomrule
            \end{tabularx}
            \end{adjustwidth}
            \noindent{\footnotesize{\textsuperscript{1} Accessed on 25 November 2022.}}
        \end{table}
        
        \subsubsection{Production Stage}
            {The catheter production stage consisted of the raw resource extraction, component production, and device assembly steps. We assume that all components were produced in the same facility, which is also where the catheter was assembled.}
            
            {A catheter was made up of a plug, handle, curvature, loop, and shaft, and weighed a total of 118.9 g. Plastic components were modelled following the case study~\cite{schulte2021combining}, which used a bill of materials provided by their collaborating remanufacturer, Vanguard AG. We assume that other materials were excluded because they do not significantly contribute to the catheter emissions, e.g., by the material cut-off rule presented in~\cite{penny2012greenhouse}. Several proprietary plastics were not available in the Ecoinvent v3.8 dataset~\cite{wernet2016ecoinvent} (PEI granulate and PA6). In these cases, we selected replacement materials that closely matched the manufacturing properties of the original plastics (polysulfone and polyamide, respectively). Most importantly, by ensuring a similar melting point, we expect to achieve high similarity in the overall plastic life cycle impacts on the modelled catheter. This is because the production process, undertaken at high temperatures, is one of the most energy and environmental impact-intensive stages of the material’s life (e.g., melting the plastic, pre-heating moulds and dies)~\cite{mannheim2021life}, while our replacements may influence other impact categories, they will create similar results for GHG emissions, our chosen climate impact (see Section~\ref{sec:metrics}).}
            
            {Once the components were manufactured (electricity and other input emissions are included in the material flows), electricity was required to assemble them into a completed catheter. Because this step was excluded in the original study, we calculated the assembly electricity as half of the electricity used to assemble and disassemble a used catheter during remanufacturing (see Section~\ref{sec:remanLifeCycle}).}
        
        \subsubsection{Sterilisation and Packaging Stages}\label{sec:virginSterilisationStage}
            {Before transportation to the user, all virgin manufactured catheters were sterilised with ethylene oxide gas. According to \cite{schulte2021combining}, the material inputs given in Table~\ref{tab:virginMaterials} were the main identifiable components from a safety sheet provided to the authors by Vanguard AG, a medical remanufacturer based in Germany. The data was also supported by AMDR's primary data. Apart from the gas materials, electricity was also needed to complete the procedure (e.g., gas introduction and evacuation from the sterilisation chamber).}

            {We assume that the packaging stage is carried out manually. Therefore, only the packaging materials are required as input to the model.}
        
        \subsubsection{Transportation Stage}
            {After production, sterilisation, and packaging, the catheters were transported from the remanufacturer in the USA to an NHS end user in the UK. There were two transportation stages: A representative 18760km container ship route between the USA and UK, and a 250km lorry transport for the final distribution within the UK. The locations were updated from the case study, which used locations specific to their collaborators (USA and Germany, respectively).}
        
        \subsubsection{Use Stage}\label{sec:virginUseStage}
            {A major update in our approach compared to \cite{schulte2021combining} is the inclusion of emissions generated during the use stage. As shown in Table~\ref{tab:includedFlows}, NHS GHG modelling standards~\cite{penny2012greenhouse} dictate that materials consumed by a medical device during the use stage should be included in the life cycle assessment. Electrophysiology catheters require only electricity to be operational. Therefore, we included a catheter's average energy use. The total amount was calculated from the average length of an electrophysiology procedure (1-hour length, primary data from AMDR) and the average watts a catheter draws~\cite{fda2007watts}. Because electrophysiology catheters are single-use devices, they do not need regular maintenance before being used.}

        \subsubsection{Incineration Stage}
            {Medical devices are most commonly incinerated after disposal to reduce contamination risk. For the end-of-life incineration after one use, we assume that the waste management facility was in the vicinity of the user, and therefore exclude transportation to the incineration site.}

    \subsection{Remanufactured Catheter}\label{sec:remanLifeCycle}
\begin{figure}[H]
            \includegraphics[width=0.99\textwidth]{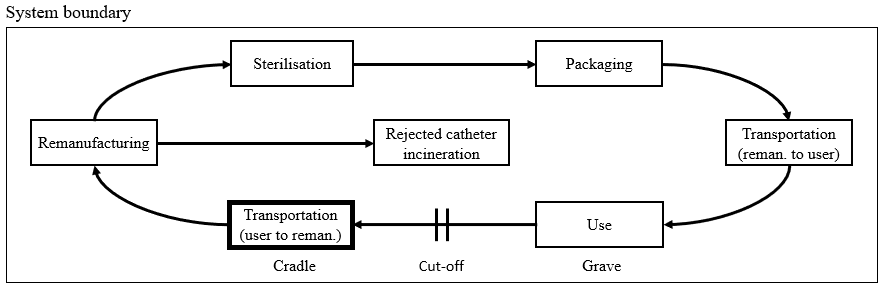}
            \caption{\emph{Remanufactured catheter product system.} The life stages are arranged in a circular life cycle from cradle-to-grave, which here is defined as from the transportation of a used catheter to its re-use after successful remanufacturing. We follow the cut-off system that is marked by the square system boundary.}\label{fig:remanLifeCycle}
        \end{figure}

        Compared to the virgin catheter, a remanufactured catheter has a circular life cycle as shown in Figure~\ref{fig:remanLifeCycle}. That starting life stage (cradle) is when a used catheter is transported from the previous user to the remanufacturer. The life cycle ends (grave) after a successfully remanufactured catheter is used by a healthcare provider.
            
        {The total weight of a remanufactured catheter is 118.9 g, the same as a virgin catheter. Table~\ref{tab:remanMaterials} presents an overview of the material flows per life stage.}
    
        \begin{table}[H]
            \caption{{\emph{Remanufacturing life stages and materials}. The materials are based on~\cite{schulte2021combining} and supplemented with primary data from our collaborators, medical device remanufacturers AMDR and Innovative Health. The data quality was assessed with the rubric in Table~\ref{tab:dataQuality} and region marks the Ecoinvent production geography.}}\label{tab:remanMaterials}
            \begin{adjustwidth}{-\extralength}{0cm}
            \begin{tabularx}{\fulllength}{lrccccX}
                \toprule
                {\textbf{Stage/Material}} & {\textbf{Quantity}} & {\textbf{Source}} & {\textbf{Quality}} & {\textbf{Time Range}} & {\textbf{Region}} & {\textbf{Description}} \\
                \midrule
                {\textbf{Transport $\times 2$}}
                 &  &  &  &  &  & {\textbf{Transport from remanufacturer to user and back.}} \\
                \multirow{2}{*}{{container   ship}} & {0.11890 kg} & \multirow{2}{*}{\href{http://ports.com}{Ports.com}\textsuperscript{1}}
                & \multirow{2}{*}{{Fair}} & \multirow{2}{*}{{2007--2021}} & \multirow{2}{*}{{GLO}} & \multirow{2}{*}{{USA-UK sea   route.}} \\
                 & {$\times$ 18,760 km}
                  &  &  &  &  &  \\
                \multirow{2}{*}{{lorry}} & {0.11890 kg} & \multirow{2}{*}{\href{https://www.google.co.uk/maps}{Google maps}\textsuperscript{1}} & \multirow{2}{*}{{Fair}} & \multirow{2}{*}{{2011--2021}} & \multirow{2}{*}{{RER}} & \multirow{2}{*}{{UK final distribution.}} \\
                 & {$\times$ 250 km} &  &  &  &  &  \\
                 \midrule
                {\textbf{Reman.}} &  &  &  &  &  & {\textbf{Dissassembly, reman., reassembly, and testing.}} \\
                {hydrogen peroxide} & {0.03020 kg} & {\cite{schulte2021combining}} & {Good} & {2018--2021} & {RER} & {Detergent ingredient.} \\
                {sodium bicarbonate} & {$1.46 \times 10^2$ kg}
                 & {\cite{schulte2021combining}} & {Good} & {2011--2021} & {GLO} & {Detergent ingredient.} \\
                {sodium cumensulph}. & {$3.50 \times 10^4$ kg} & {\cite{schulte2021combining}} & {Good} & {2015--2021} & {GLO} & {Detergent ingredient}. \\
                {tap water} & {7.00000 kg} & {\cite{schulte2021combining}} & {Good} & {2012--2021} & {Europe} & {Process water.} \\
                {water, ultrapure} & {5.00000 kg} & {Adjusted~\cite{schulte2021combining}} & {Fair} & {2009--2021} & {CA-QC} & {Process water.} \\
                {electricity} & {0.207 kWh} & {\cite{schulte2021combining}} & {Good} & {2014--2021} & {US-WECC} & {Process electricity.} \\
                \midrule
                {\textbf{Incineration}} &  &  &  &  &  & {\textbf{End-of-life waste disposal.}} \\
                {plastic incineration} & {0.01784 kg} & {AMDR, IH} & {Good} & {2006--2021} & {RoW} & {Municipal (15\% rejection rate).} \\
                \midrule
                {\textbf{Sterilisation}} &  &  &  &  &  & {\textbf{Gas sterilisation.}} \\
                {carbon dioxide} & {$2.82 \times 10^3$ kg} & {\cite{schulte2021combining}} & {Good} & {2011--2021} & {RER} & {Sterilisation gas ingredient.} \\
                {ethylene oxide} & {$1.80 \times 10^4$ kg} & {\cite{schulte2021combining}} & {Good} & {2018--2021} & {RER} & {Sterilisation gas ingredient.} \\
                {electricity} & {0.360 kWh} & {Adjusted~\cite{schulte2021combining}} & {Good} & {2014--2021} & {US-WECC} & {Process electricity.} \\
                \midrule
                {\textbf{Packaging}} &  &  &  &  &  & {\textbf{Manual packaging.}} \\
                {polyethylene HD} & $2.00 \times 10^2$ kg & {\cite{schulte2021combining}} & {Good} & {2011--2021} & {GLO} & {Primary (sterile) packaging.} \\
                {carton box} & {$1.40 \times 10^1$ kg} & {Adjusted~\cite{schulte2021combining}} & {Fair} & {2011--2021} & {GLO} & {Secondary packaging.} \\
                \midrule
                {\textbf{Use}} &  &  &  &  &  & {\textbf{Materials consumed by device.}} \\
                {electricity} & {0.050 kWh} & {AMDR, \cite{fda2007watts}} & {Very Good} & {2014--2021} & {GB} & {1 h of use, 50 watts.} \\
                \bottomrule
            \end{tabularx}
            \end{adjustwidth}
            \noindent{\footnotesize{\textsuperscript{1} Accessed on 25 November 2022.}}
        \end{table}
        
        \subsubsection{Transportation Stage}
            {In a significant change from \cite{schulte2021combining}, we have doubled the transportation route between the USA and the UK. The case study assumes that the used catheters and remanufactured catheters are transported in a single round trip. However, discussions with medical device remanufacturers AMDR and Innovative Health revealed that this does not reflect current industry practice. Each transport route has two stages: A representative 18,760 km container ship route between the USA and UK, and a 250 km lorry transport for the final distribution within the UK.}
        
        \subsubsection{Remanufacturing Stage}
            {The catheter remanufacturing stage consists of the disassembly, cleaning, reassembly, and testing steps. We assume that all processes take place in the same remanufacturing facility, and therefore do not include further transportation. Additionally, we assume that the electrophysiology catheter's software does not have to be maintained. Used catheters that failed quality control at any point of the process or passed their cleared number of maximum turns reached their end of life. For simplicity, the modelled remanufacturing process does not include replacing faulty components.}
            
            {A significant difference of our model to~\cite{schulte2021combining} is the rejection rate. The case study assumes an almost 50\% rejection rate. In other words, only one catheter is successfully reprocessed for every two used catheters that arrive at the remanufacturing facility. In comparison, medical device remanufacturers AMDR and Innovative Health confirmed that a 15\% rejection rate more accurately reflects the current state of EP catheter remanufacturing in 2021.}
        
        \subsubsection{Incineration Stage}
            {As with the virgin catheter, we assume that incineration occurs near the healthcare provider and therefore exclude transportation to the waste management facility (see \linebreak Section~\ref{sec:virginLifeCycle}).}

        \subsubsection{Sterilisation and Packaging Stages}
            The gas sterilisation and packaging processes are assumed to be exactly the same as for a virgin single-use catheter (see Section~\ref{sec:virginSterilisationStage}).
        
        \subsubsection{Use Stage}
            According to NHS regulations~\cite{standard2013ghg}, medical device remanufacturing must return a used catheter to its original state to classify it as a remanufactured single-use device. Consequently, we may assume that the use stage of a fully remanufactured catheter is the same as that of a virgin manufactured catheter (see Section~\ref{sec:virginUseStage}).

\section{Greenhouse Gas Emission Results}\label{sec:ghgResults}
    This section contains our comparative analysis of virgin manufactured and remanufactured catheter emissions. All results were generated following the comprehensive methodology illustrated in Section~\ref{sec:methodology}. Wider-context insights and implications initiated by the results are discussed in Section~\ref{sec:discussion}.
    
   {The functional unit of analysis is defined as the production and potential disposal of one catheter after its use in a 1-h procedure. We report the following values:}
    \begin{enumerate}[label=(\roman*)]
        \item {Burden-free per turn emissions $E^f_v$ for virgin manufactured and $E^f_r$ for remanufactured catheters;}
        \item {Burdened per turn catheter emissions $E^b_{perTurn}$;}
        \item {Furthermore, burdened per life catheter emissions $E^b_{perLife}$.}
    \end{enumerate}
    
    {Readers are referred to Section~\ref{sec:metrics} for a more detailed treatment of the metrics with definitions, equations, and intuitions.}

    \subsection{Burden-Free Catheter Emissions}\label{sec:burdenfreeResults}
\begin{figure}[H]
            \begin{subfigure}{0.49\textwidth}
                \centering
                \includegraphics[width=\textwidth]{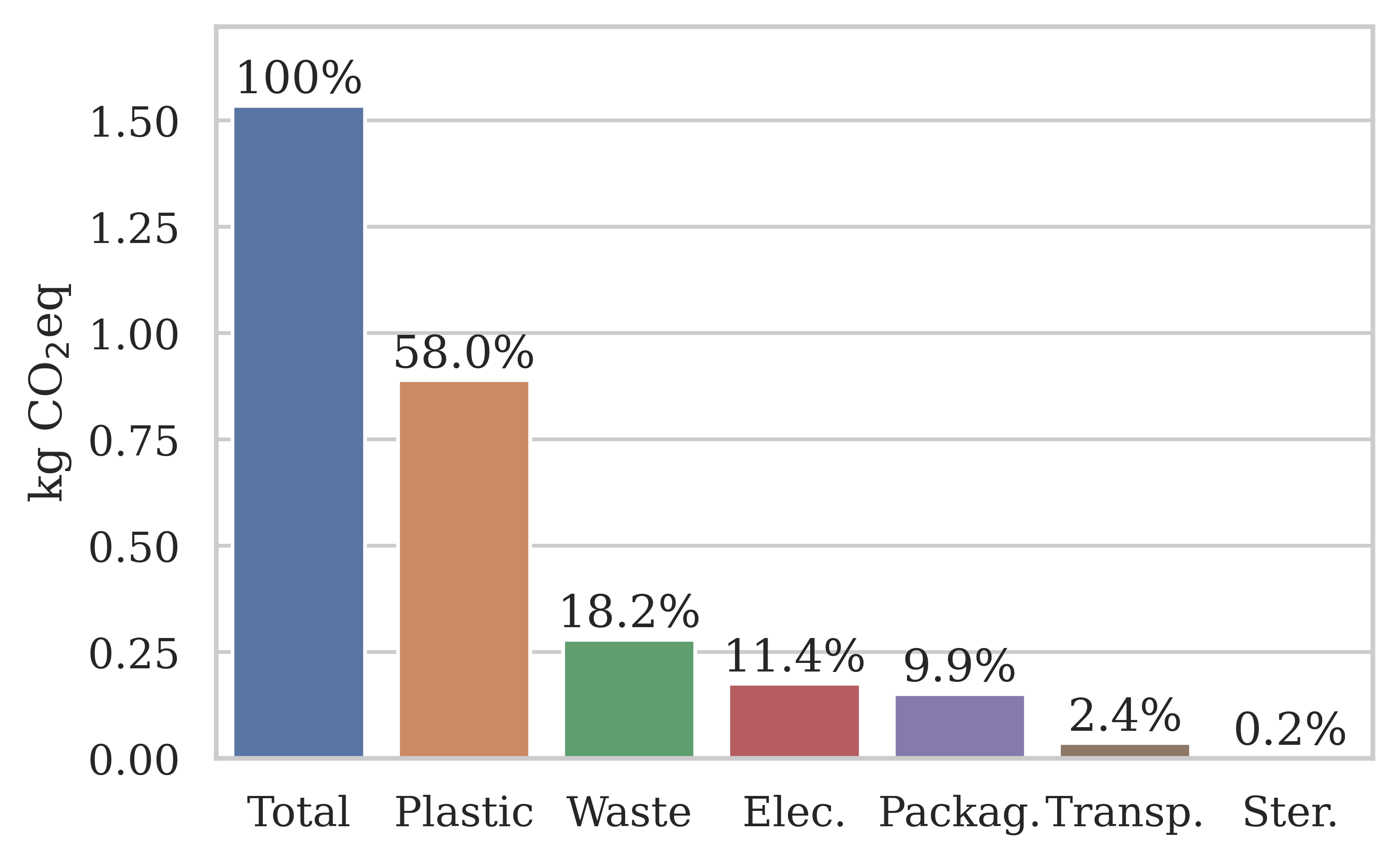}
                \captionsetup{position=bottom,justification=centering}
                \caption{{}}\label{fig:virginEmissionBreakdown}
            \end{subfigure}
            \hfill
            \begin{subfigure}{0.49\textwidth}
                \centering
                \includegraphics[width=\textwidth]{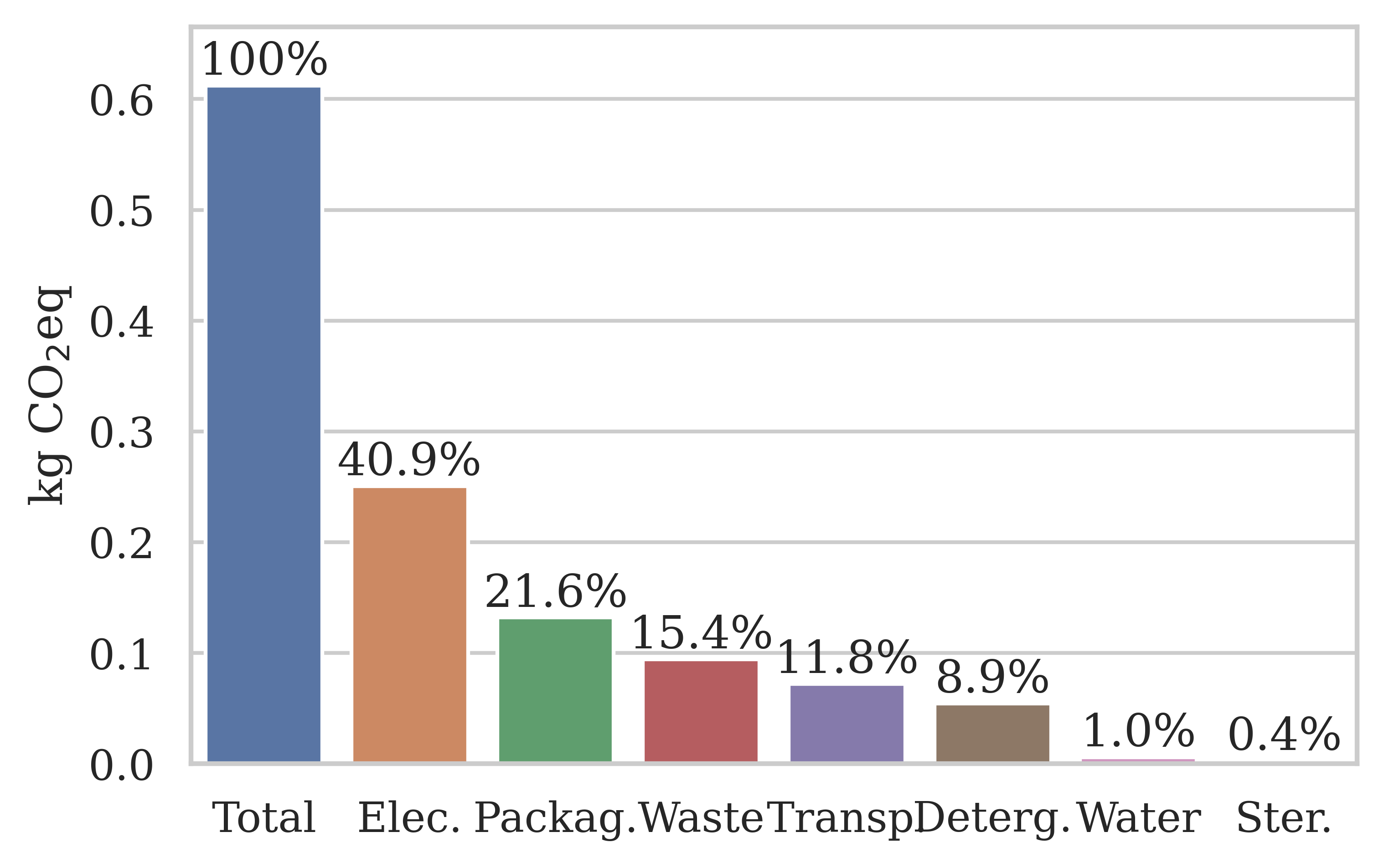}
                \captionsetup{position=bottom,justification=centering}
                \caption{{}}
                \label{fig:remanEmissionBreakdown}
            \end{subfigure}
            \caption{{\emph{Burden-free catheter results per turn.} Virgin emissions are primarily caused by plastic processing and incineration. In contrast, remanufactured emissions are more evenly distribution across categories. (\textbf{a}) Virgin catheter emissions} $E^f_v$. (\textbf{b}) Remanufactured catheter emissions $E^f_r$.}
            \label{fig:catheterEmissionsBreakdown}
        \end{figure}

        {The total climate emissions generated during the production, use, and disposal of one \emph{virgin catheter} is $\bm{E^f_v = 1.53}$\textbf{kg CO$\bm{_2}$eq} (based on the material inputs shown in Table~\ref{tab:virginMaterials}). Figure~\ref{fig:catheterEmissionsBreakdown}a shows how the total value is broken down into material categories. For a different perspective, interested readers are referred to Table~\ref{tab:materialEmissions}a for the emissions per life stage and individual material. As expected, the procurement and manufacturing of plastic catheter components were responsible for the majority of virgin catheter emissions, coming in at 58\%. Disposing of the waste plastic was also a significant contributor, covering 18\%. In total, around three-quarters of virgin catheter emissions came from the production and disposal of plastic components, which implies that remanufacturing may make a consequential dent in life cycle emissions. The two lowest contributors were gas sterilisation (0.2\%) and, surprisingly, transportation of catheters between the USA-based manufacturer and the UK-based user (2.4\%). We will explore this figure in more detail as part of the sensitivity analysis in Section~\ref{sec:sensitivtyRemanLocation}.}
        
        \vspace{-6pt}
        \begin{table}[H]
       \setlength{\tabcolsep}{4.56mm}

            \caption{{\emph{Greenhouse gas emissions by life stage and material.} Around three-quarters of virgin catheter emissions came form the production and disposal of plastic components. In comparison, remanufacturing emissions were more evenly divided across life stages.}}\label{tab:materialEmissions}

                 \begin{adjustwidth}{-\extralength}{0cm}
                \begin{tabularx}{\fulllength}{XrrXrr	}
                    \toprule
                 \multicolumn{3}{c}{\textbf{(a) Virgin manufactured catheter}} & \multicolumn{3}{c}{\textbf{(b) Remanufactured catheter}}   \\

                     \cmidrule(lr){1-3}\cmidrule(lr){4-6}
                    {\textbf{Stage/Material}} & {\textbf{Kg CO$_2$eq}} & {\textbf{Proportion}} &{\textbf{Stage/Material}} & {\textbf{Kg CO$_2$eq}} & {\textbf{Proportion}}\\
                    \cmidrule(lr){1-3}\cmidrule(lr){4-6}
                    {\textbf{Production}} & \multicolumn{1}{r}{{\textbf{0.90018}}} & \multicolumn{1}{r}{{\textbf{58.65\%}}} &  {\textbf{Transport  $\times$2}} & \multicolumn{1}{r}{{\textbf{0.07223}}} & \multicolumn{1}{r}{{\textbf{11.80\%}}} \\

                    {polyamide} & {0.02934} & {1.91\%} & {container ship} & \multicolumn{1}{r}{{0.04203}} & {6.87\%}\\
                    {ethylene   glycol} & {0.00261} & {0.17\%}& {lorry} & \multicolumn{1}{r}{{0.03020}} & {4.93\%}\\
                    \cmidrule(lr){4-6}
                    {polyethylene   LD} & {0.00077} & {0.05\%} & {\textbf{Remanufacturing}} & \multicolumn{1}{r}{{\textbf{0.14603}}} & \multicolumn{1}{r}{{\textbf{23.85\%}}}\\
                    {polysulfone} & {0.85250} & {55.55\%}& {hydrogen peroxide} & \multicolumn{1}{r}{{0.03497}} & {5.71\%}\\
                    {polyurethane} & {0.00419} & {0.27\%}& {sodium bicarbonate} & \multicolumn{1}{r}{{0.01861}} & {3.04\%}\\
                    {electricity} & {0.01077} & {0.70\%} & {sodium cumensulph.} & \multicolumn{1}{r}{{0.00084}} & {0.14\%}\\
                    \cmidrule(lr){1-3}
                    {\textbf{Sterilisation}} & \multicolumn{1}{r}{{\textbf{0.15171}}} & \multicolumn{1}{r}{{\textbf{9.89\%}}}&{tap water} & \multicolumn{1}{r}{{0.00232}} & {0.38\%} \\
                    {carbon   dioxide} & {0.00229} & {0.15\%}&{water, ultrapure} & \multicolumn{1}{r}{{0.00368}} & {0.60\%} \\
                    {ethylene   oxide} & {0.00034} & {0.02\%}&{electricity} & \multicolumn{1}{r}{{0.08561}} & \multicolumn{1}{r}{{13.98\%}} \\
                     \cmidrule(lr){4-6}
                    {electricity} & \multicolumn{1}{r}{{0.14908}} & {9.72\%} &  {\textbf{Incineration}} & \multicolumn{1}{r}{{\textbf{0.09426}}} & \multicolumn{1}{r}{{\textbf{15.40\%}}}\\
                     \cmidrule(lr){1-3}
                    {\textbf{Packaging}} & \multicolumn{1}{r}{{\textbf{0.15188}}} & \multicolumn{1}{r}{{\textbf{9.90\%}}} &{plastic incineration} & \multicolumn{1}{r}{{0.09426}} & {15.40\%}\\
                      \cmidrule(lr){4-6}
                    {polyethylene   HD} & {0.04798} & {3.13\%}&{\textbf{Sterilisation}} & \multicolumn{1}{r}{{\textbf{0.15182}}} & \multicolumn{1}{r}{{\textbf{24.80\%}}} \\
                    {carton   box} & {0.10390} & {6.77\%}&{carbon dioxide} & \multicolumn{1}{r}{{0.00229}} & {0.37\%} \\
                    \cmidrule(lr){1-3}
                    {\textbf{Transport}} & \multicolumn{1}{r}{{\textbf{0.03612}}} & \multicolumn{1}{r}{{\textbf{2.35\%}}} & {ethylene oxide} & \multicolumn{1}{r}{{0.00034}} & {0.06\%}\\
                    {container shi}p & {0.02102} & {1.37\%} &{electricity} & \multicolumn{1}{r}{{0.14919}} & \multicolumn{1}{r}{{25.24\%}}\\
                      \cmidrule(lr){4-6}
                    {lorry} & 0.{01510} & {0.98\%} &  {\textbf{Packaging}} & \multicolumn{1}{r}{\textbf{{0.13201}}} & \multicolumn{1}{r}{{\textbf{21.57\%}}}\\
                    \cmidrule(lr){1-3}
                    {\textbf{USE}} & \multicolumn{1}{r}{{\textbf{0.01586}}} & \multicolumn{1}{r}{{\textbf{1.04\%}}} & {polyethylene HD} & \multicolumn{1}{r}{{0.04798}} & {7.84\%}\\
                    {electricity} & {0.01586} & \multicolumn{1}{r}{{1.04\%}} & {carton box} & \multicolumn{1}{r}{{0.08403}} & {13.73\%}\\
                    \cmidrule(lr){1-3}\cmidrule(lr){4-6}
                    {\textbf{Incineration}} & \multicolumn{1}{r}{{\textbf{0.27902}}} & {18.18\%} & {\textbf{Use}} & \multicolumn{1}{r}{{\textbf{0.01586}}} & \multicolumn{1}{r}{{\textbf{2.59\%}}}\\
                    {plastic   incineration} & {0.27902} & {18.18\%} &{electricity} & \multicolumn{1}{r}{{0.01586}} & {2.59\%}\\
                    \cmidrule(lr){1-3}\cmidrule(lr){4-6}
                    {$\bm{E_v^f}$} & \multicolumn{1}{r}{{1.53477}} & \multicolumn{1}{r}{{100.00\%}} &{$\bm{E_r^f}$} & \multicolumn{1}{r}{{0.61221}} & \multicolumn{1}{r}{{100.00\%}}\\
                    \bottomrule
                \end{tabularx}

            \end{adjustwidth}
        \end{table}
        
        {Turning to a \emph{remanufactured catheter}, climate emissions totalled $\bm{E^f_r = 0.61}$\textbf{kg CO$\bm{_2}$eq}} (Note that the used catheter at the start of the remanufacturing process had no emissions tied to it, as dictated by the cut-off model we used (see Section~\ref{sec:methodology})).
{(based on the material inputs shown in Table~\ref{tab:remanMaterials}). Figure~\ref{fig:catheterEmissionsBreakdown}b illustrates the emission breakdown into categories. For more detail on the life stage and material emissions, see Table~\ref{tab:materialEmissions}b. Compared to the virgin catheter, emissions were more evenly divided across categories. Electricity was the highest single contributor (41\%), which suggests that the composition of energy production may be a significant factor when optimising the remanufacturing process. The lowest climate impact categories were water (1\%) and sterilisation gas (0.4\%).}
        
        {\emph{Comparing a virgin catheter to a remanufactured catheter}, we found that \textbf{remanufacturing reduced climate emissions by 60\%} per burden-free catheter ($-$1.92 kg CO$_2$eq, Table~\ref{tab:ourResultsBurdenfree}). As expected, the single largest emission saving came from the omission of plastic processing and component manufacture ($-$1.89 kg CO$_2$eq) with reduced waste processing in second place ($-$1.19 kg CO$_2$eq). In contrast, the inclusion of cleaning inputs for remanufacturing (detergent, water, electricity) only marginally increased the emissions by +0.14 kg CO$_2$eq. A final noteworthy change is that the transportation emissions were doubled for the remanufactured catheter, since the used catheters had to be transported to the remanufacturing facility and back to the user. However, total transportation emissions were still nominal (0.07 kg CO$_2$eq).}
        
        \begin{table}[H]
           \setlength{\tabcolsep}{10.56mm}

            \caption{{\emph{Burden-free catheter remanufacturing emission savings.} Our analysis shows that \textbf{catheter remanufacturing reduces climate emissions by 60\%}}. The results are given in kg CO$_2$eq.}\label{tab:ourResultsBurdenfree}

            \begin{tabularx}{\textwidth}{lrrr}
                \toprule
                {\textbf{Category}} & \multicolumn{1}{c}{{\textbf{Virgin}}} & \multicolumn{1}{c}{{\textbf{Reman.}}} & \multicolumn{1}{c}{{\textbf{$\Delta$}} }\\
                \cmidrule(lr){1-2}\cmidrule(lr){3-4}
                {Plastic} & {0.889} & {0} & {$-$1.889} \\
                {Waste} & {0.279} & {0.094} & {$-$1.185} \\
                {Elec.} & {0.176} & {0.251} & {+0.075} \\
                {Packag.} & {0.152} & {0.132} & {$-$1.020} \\
                {Transp.} & {0.036} & {0.072} & {+0.036} \\
                {Deterg.} & {0} & {0.054} & {+0.054} \\
                {Water} & {0} & {0.006} & {+0.006} \\
                {Ster.} & {0.003} & {0.003} & {0} \\
                \cmidrule(lr){1-2}\cmidrule(lr){3-4}
                {\textbf{$\sum$}} & {1.535} & {\textbf{0.612}} & {\textbf{$-$1.923}} \\
                \bottomrule
            \end{tabularx}
        \end{table}

    \subsection{Sensitivity Analysis and Burdened Catheter Emissions}\label{sec:sensitivityAnalysis}
       {In this section, we carry out a comprehensive sensitivity analysis of three high-impact remanufacturing life cycle parameters:}
        \begin{enumerate}[label=(\roman*)]
            \item {The \emph{remanufacturing location} $L$.}
            \item {The \emph{rejection rate} $R$.}
            \item {Furthermore, the \emph{number of turns} $N$.}
        \end{enumerate}
        
        {After examining the magnitude of each key parameter's individual effect on emissions (Sections~\ref{sec:sensitivtyRemanLocation}--\ref{sec:sensitivityNrTurns}), we contextualise the results by analysing three representative scenarios (Section~\ref{sec:sensitivityScenarios}). Readers are referred to Section~\ref{sec:sensitivityParameters} for a detailed description of the methodology and parameters.}
        
        \subsubsection{Remanufacturing Location}\label{sec:sensitivtyRemanLocation}
            Remanufacturing is commonly offered as a third-party service~\cite{opresnik2015manufacturer}. When comparing catheter life cycle emissions, we must consider both the absolute and relative location of the remanufacturing facility to the user because both may have wide-ranging effects: from the transport mode and distance to the origin and composition of the remanufacturing materials and processes (e.g., electricity generation, water processing, waste treatment).
            
\begin{figure}[H]
                \includegraphics[width=0.75\textwidth]{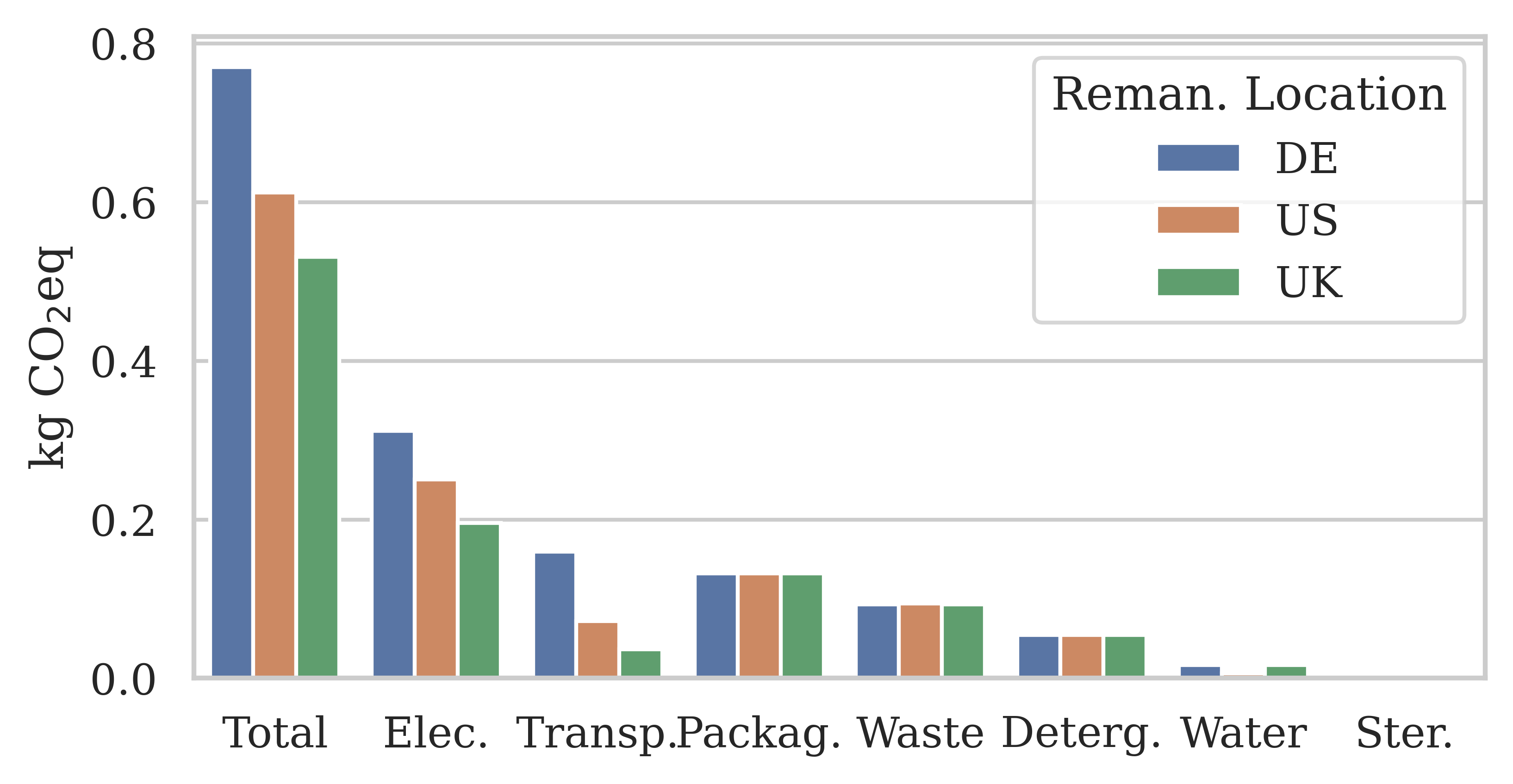}
                \caption{{\emph{Sensitivity analysis of the remanufacturing location $L$.} Burden-free emission results $E^f_r$ show that the location may influence non-transport categories because the regionality of their material inputs has changed.}}\label{fig:sensitivityRemanLocation}
            \end{figure}

            {Figure~\ref{fig:sensitivityRemanLocation} shows the burden-free emission results $E^f_r$ for one catheter remanufactured in three different locations: Germany (0.77 kg CO$_2$eq), USA (0.61 kg CO$_2$eq), and UK \linebreak (0.53 kg CO$_2$eq). In each scenario, the remanufactured catheter was used in the UK. \emph{Electricity} showed a noticeable emissions increase when remanufacturing took place in Germany (+0.06 kg CO$_2$eq), potentially because energy generation is more reliant on coal combustion than other countries~\cite{bruninx2013impact}. Conversely, UK remanufacturing reduced electricity emissions by $-$1.6 kg CO$_2$eq. Since electricity was the highest emission category, the composition of energy production technologies should be taken into consideration to optimise the remanufacturing process.}
            
            Surprisingly, the \emph{transportation} of used and remanufactured catheters between Germany and the UK had the highest climate impact as well (0.16 kg CO$_2$eq), over USA-UK transportation (0.07 kg CO$_2$eq). This may be partially due to the different transport modes. Our results suggest that in terms of their climate impact, long-distance routes using primarily container ship transportation (USA-UK, 19,000 km) may be more efficient than shorter-distance routes using primarily road transportation (DE-UK, 1400 km). Interestingly, the total transport distance may not necessarily indicate the emission efficiency of a remanufacturing location.
            
            Given its significance to the majority of remanufacturing materials and processes, it is unexpected to see multiple categories with only minor changes between the burden-free emissions at three different locations. We hypothesise that this is most likely due to two~reasons:
            \begin{enumerate}[label=(\roman*)]
                    \item The materials (e.g., for cleaning and sterilisation) were in small quantities, and therefore the absolute regional differences were minimal,
                    \item And/or Ecoinvent v3.8~\cite{wernet2016ecoinvent} had only broad regional categories that included multiple remanufacturing locations such as the `global' region for waste treatment, packaging, and water.
                \end{enumerate}
                
            For those categories, more detailed and regionally distinguished data flows could be used in future for more precise emission calculations.
            
            Even with the stated regionality limitations, our results highlight the environmental impact a catheter user may exert through the careful selection of a remanufacturer based on their facility's location. Our results suggest that the location of remanufacturing may not just have a significant effect on the emissions caused by the transportation, but also on the climate impact of the remanufacturing process itself.

        \subsubsection{Rejection Rate}\label{sec:sensitivityRejection}
            The rejection rate describes the proportion of used catheters that are discarded due to sub-par quality for every successfully remanufactured catheter. Many factors, including the catheter model and collection processes, may influence the quality rejection rate. For simplicity, we assume that low-quality catheters are rejected on arrival at the remanufacturing facility (e.g., because of missing identification or damage during collection).
            
\begin{figure}[H]
                \includegraphics[width=0.75\textwidth]{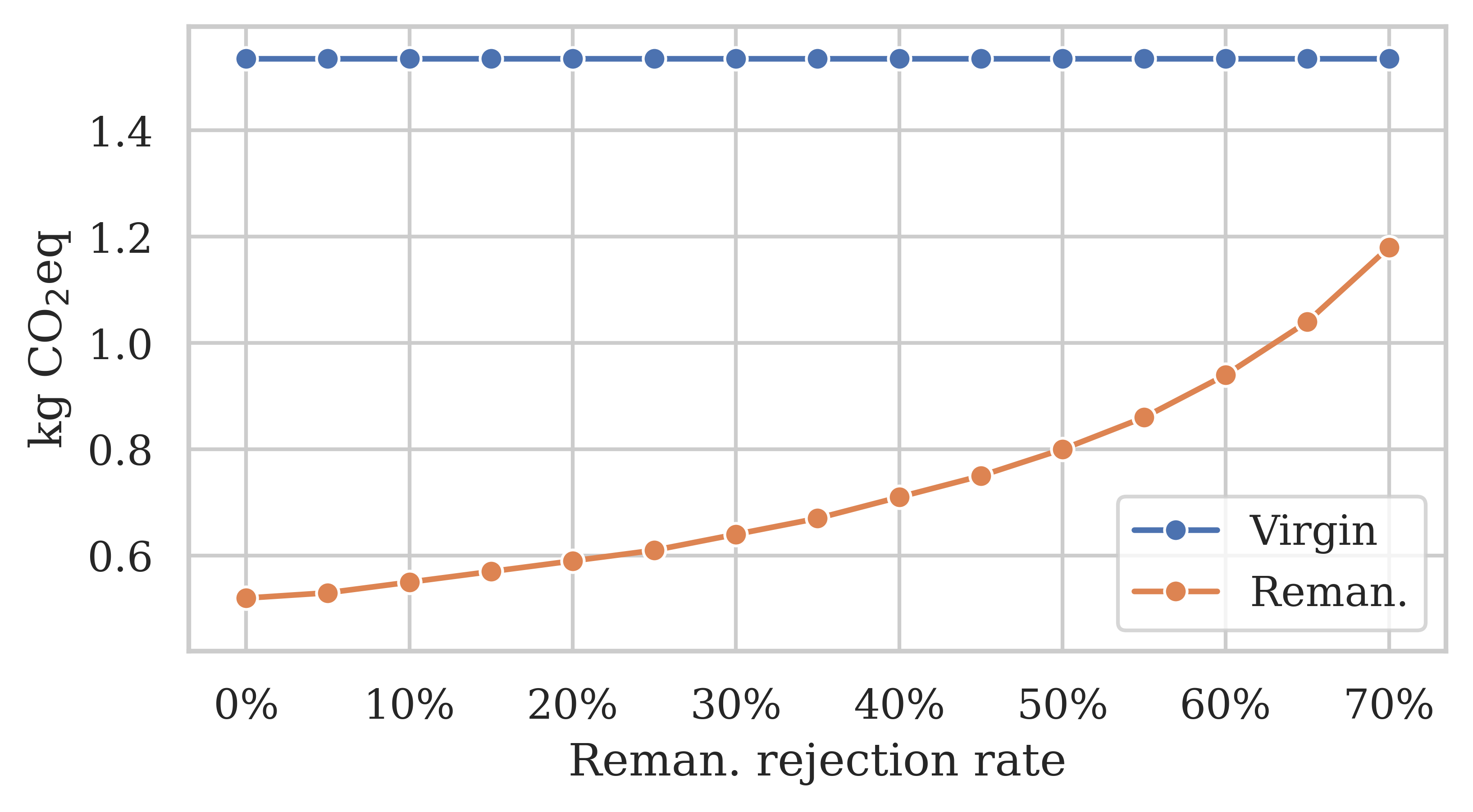}
                \caption{{\emph{Sensitivity analysis of the rejection rate $R$.} Burden-free remanufacturing emissions $E^f_r$ increase exponentially as the rejection rate grows. Since 100\% of virgin catheters are discarded after use, their emissions $E^f_v$ are not affected.}}\label{fig:sensitivtyRejectionRate}
            \end{figure}

            {To illustrate the significant influence of the rejection rate on remanufacturing emissions, we evaluated a range from 0\% (optimal) to 70\% in Figure~\ref{fig:sensitivtyRejectionRate}. The virgin catheter emissions are given for reference ($E^f_r = 1.53$kg CO$_2$eq). As the rejection rate increases, the burden-free remanufactured catheter emissions $E^f_r$ increased exponentially from 0.52 to 1.18 kg CO$_2$eq. To put that in context relative to a virgin catheter, a 0\% rejection rate represents a 66\% emission reduction through remanufacturing, 35\% represents 56\% reduction, and 70\% means only a 23\% reduction, respectively.}
            
            The increased emission savings were directly caused by a lower quality rejection rate, which implies that more of the used catheters arriving at the remanufacturing facility could be successfully reprocessed. In turn, this led to less used catheters reaching their end of life due to a low quality. The climate impact reduction primarily stemmed from the reduced waste incineration per reprocessed catheter.

        \subsubsection{Number of Turns}\label{sec:sensitivityNrTurns}
\begin{figure}[H]
                \includegraphics[width=0.6\textwidth]{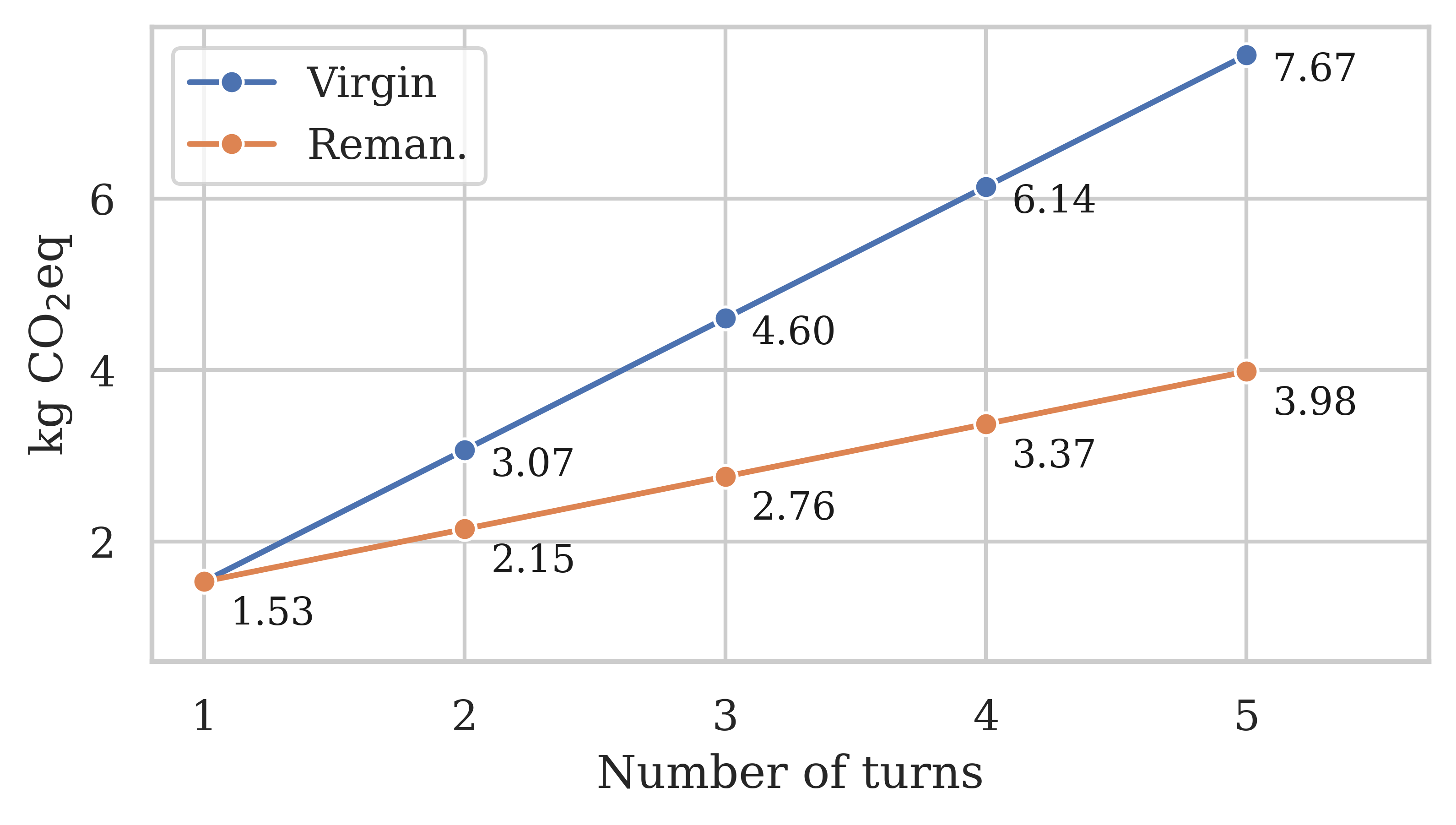}
                \caption{{\emph{Sensitivity analysis of the number of turns $N$.} Each remanufactured catheter has its first life as a virgin catheter with emission $E^f_v$. After that, remanufactured burdened per-life emissions $E^b_{perLife}$ increase by $E^f_r$ per turn.}}\label{fig:sensitivityNrTurns}
            \end{figure}

            To ensure that the functionality of remanufactured catheters is substantially equivalent to a virgin catheter, each catheter model must be cleared for a certain number of turns by regulators before they may be remanufactured. According to medical device remanufacturer Innovative Health, EP catheters are relatively fragile compared to other remanufactured devices and typically cannot withstand more than 1--5 turns. Consequently, we evaluated the burdened emissions $E^b_{perLife}$ of remanufactured catheters for that range and contrasted the results against the total emissions of using a new virgin catheter for each turn in Figure~\ref{fig:sensitivityNrTurns}.

            {Both scenarios started at $E^f_v = 1.53$kg CO$_2$eq since a remanufactured catheter's first life is as a virgin catheter. As the number of turns increased, so did the improved savings of remanufactured over virgin catheters. In the virgin scenario, each `turn' required the production, use, and disposal of a brand-new catheter, which meant a linear growth of \linebreak +1.53 kg CO$_2$eq. In contrast, remanufacturing the same catheter repeatedly had a significantly shallower growth of +0.61 kg CO$_2$eq per turn. This represents a 60\% reduction per turn since a remanufactured catheter only emits 40\% of emissions compared to a virgin catheter.}
            
            {Our results highlight that remanufacturing emission savings increase significantly with the number of viable turns. After two turns, $E^b_{perLife}$ is decreased by 30\% compared to two virgin catheters ($-$0.9 kg CO$_2$eq), 40\% after three turns ($-$1.8 kg CO$_2$eq), and 48\% after five turns ($-$3.7 kg CO$_2$eq). To put it in a different perspective, a remanufactured catheter used five times has lower climate emissions than three virgin catheters (4 kg CO$_2$eq vs. \linebreak 4.6 kg CO$_2$eq).}

        \subsubsection{Multivariate Scenarios}\label{sec:sensitivityScenarios}
            Finally, we combine the individually explored parameters into three representative multivariate scenarios, as shown in Table~\ref{tab:sensitivityScenarios}. By analysing these scenarios in parallel, we may contextualise the univariate sensitivity analysis results (Sections~\ref{sec:sensitivtyRemanLocation}--\ref{sec:sensitivityNrTurns}) and evaluate the possible overall gains that may be achieved with an improved remanufacturing process.
            
            \begin{table}[H]
            \setlength{\tabcolsep}{6.56mm}

                \caption{\emph{Multivariate sensitivity analysis scenarios.} High-impact remanufacturing parameters were combined into three scenarios to contextualise emission results: remanufacturing location $L$, used catheter rejection rate $R$, and number of turns $N$.}\label{tab:sensitivityScenarios}

                \begin{tabularx}{\textwidth}{lcrrl}
                    \toprule
                    \textbf{Scenario} & \multicolumn{1}{c}{\boldmath{$L$}} & \multicolumn{1}{c}{\boldmath{$R$}} & \multicolumn{1}{c}{\boldmath{$N$}} & \textbf{Comments} \\
                    \midrule
                    \textbf{Good} & UK & 0\% & 5 & Ideal scenario. \\
                    \textbf{Average} & DE & 15\% & 4 & Approximately current. \\
                    \textbf{Bad} & USA & 50\% & 3 & Approximately following \cite{schulte2021combining}. \\
                    \bottomrule
                \end{tabularx}
            \end{table}
            
\begin{figure}[H]
                \begin{subfigure}{0.5235\textwidth}
                    \centering
                    \includegraphics[width=\textwidth]{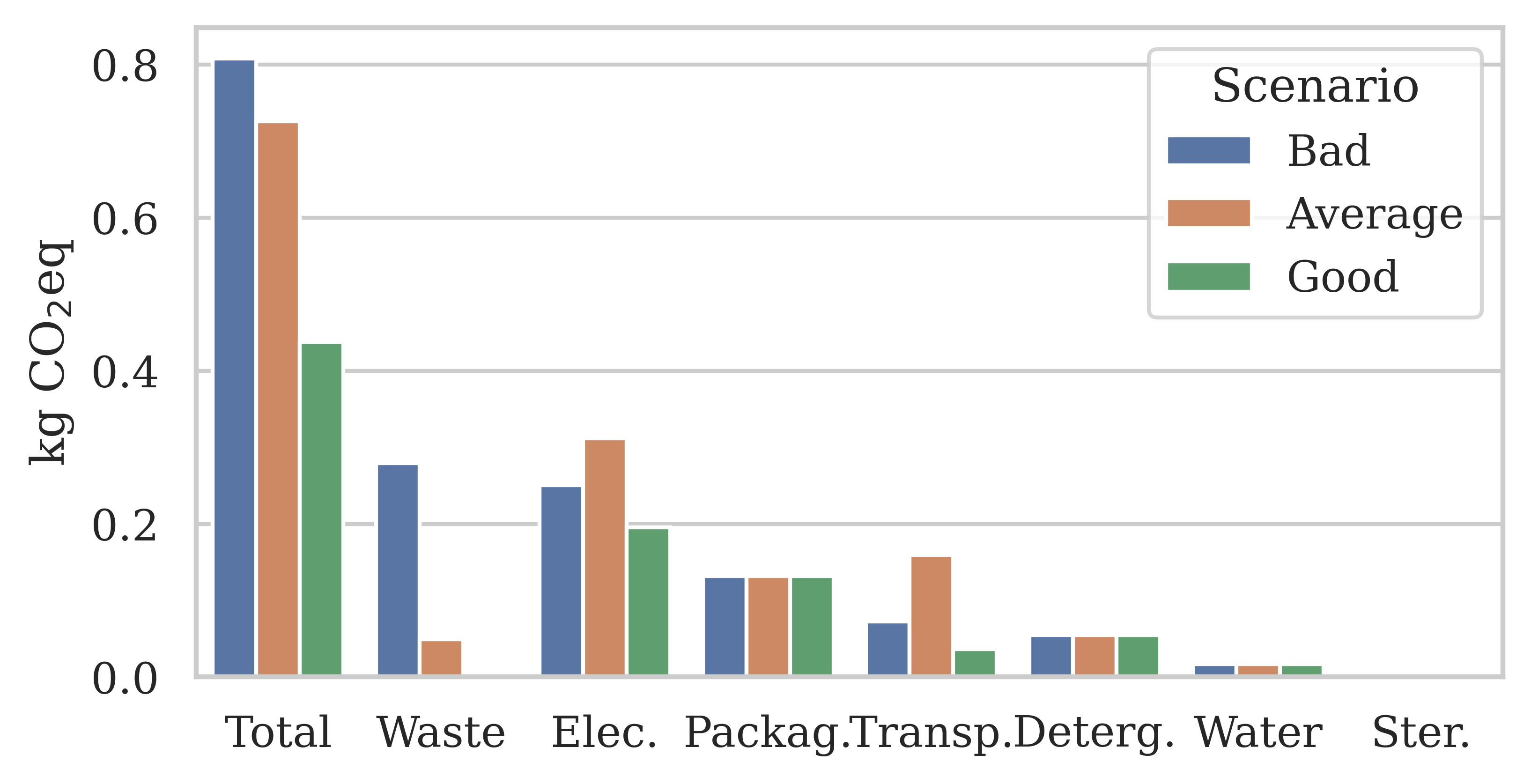}
                    \captionsetup{position=bottom,justification=centering}
                    \caption{{}}\label{fig:sensitivityBurdenFree}
                \end{subfigure}
                \hfill
                \begin{subfigure}{0.4365\textwidth}
                    \centering
                    \includegraphics[width=\textwidth]{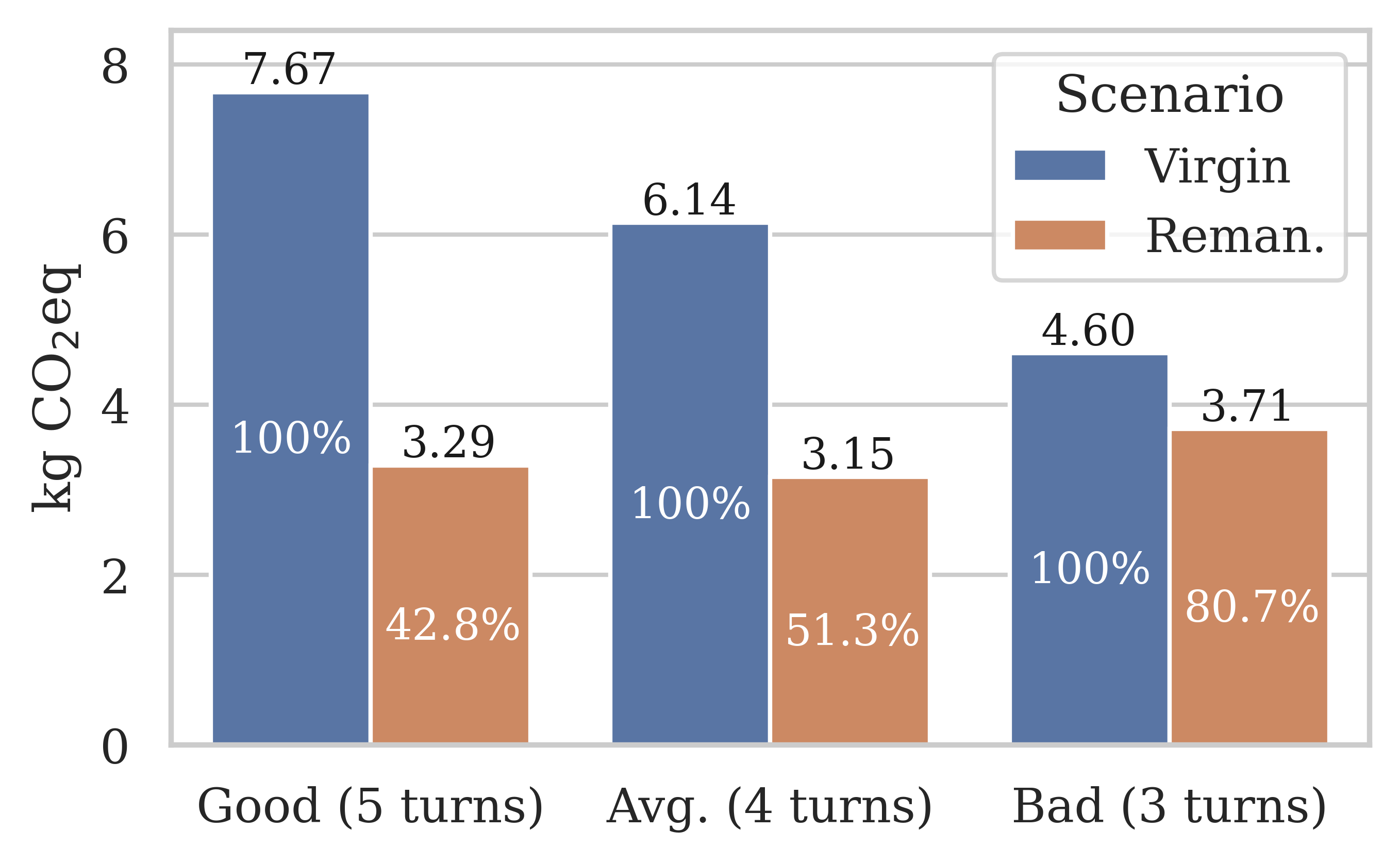}
                    \captionsetup{position=bottom,justification=centering}
                    \caption{{}}
                    \label{fig:sensitivtyPerLife}
                \end{subfigure}
                \caption{{\emph{Multivariate sensitivity results of remanufactured catheters.} As the efficiency of the remanufacturing process improves (see Table~\ref{tab:sensitivityScenarios}), the emission savings of remanufactured catheters improves compared to an equivalent number of virgin catheters ($E^f_v \cdot N$). Remanufacturing reduces emissions by up to 57\%. (\textbf{a}) Burden-free per turn emissions} $E^f_r$. (\textbf{b}) Burdened per life emissions $E^b_{perLife}$.}\label{fig:sensitivityScenarios}
            \end{figure}

            {A high-level comparison in Figure~\ref{fig:sensitivityScenarios}a revealed that the burden-free climate emissions of a remanufactured catheter $E^f_r$ followed the expected ranking (0.81, 0.73, 0.44 kg CO$_2$eq). The Average scenario improved on the Bad remanufacturing scenario by 10\% compared to the 46\% improvement by the good scenario. The largest variations in results were observed in the waste treatment (0--0.3 kg CO$_2$eq) and transport (0.04--0.16 kg CO$_2$eq) categories, which were the most directly affected by the quality rejection rate and remanufacturing location parameters. We also observed a variation in the electricity category, influenced by the remanufacturing location.}

            {To incorporate the number of turns, we calculated the burdened per life emissions $E^b_{perLife}$ for remanufactured catheters and compared them to an equal number of virgin catheters ($E^f_v \cdot N$) in Figure~\ref{fig:sensitivityScenarios}b. As the remanufacturing process became more efficient overall (Bad to Average to Good scenarios), the \textbf{per life emission reductions increased significantly from 19\% to 57\%} ($-$0.9 to $-$4.3 kg CO$_2$eq). Our results showcase that even sub-optimal remanufacturing processes with relatively small per-turn emission reductions (e.g., the Bad scenario) may significantly reduce the emissions across a remanufactured catheter's entire life cycle.}

    \subsection{Long-Term Emissions Use Case: A Realistic Buy-Back Scheme}\label{sec:buyBack}
        {Finally, we simulate an industry buy-back scheme where medical remanufacturing is provided by a third-party service. Given the number of turns $N$ and the remanufacturing rejection rate $R$, we may calculate the necessary initial injection of virgin catheters $C$. Results for catheter uses $U = 1000$ are shown in Table~\ref{tab:buyBackInjection} for the Bad, Average, and Good scenarios, calculated with Equation~\eqref{eq:catheterUses} (defined in Section~\ref{sec:longTermMetric}).}
    
        \begin{table}[H]
        \setlength{\tabcolsep}{9.56mm}

            \caption{\emph{The initial injection $C$ of virgin catheters needed to prime a buy-back scheme system}. Results are calculated with Equation~\eqref{eq:catheterUses} for the three scenarios described in Table~\ref{tab:sensitivityScenarios}.}\label{tab:buyBackInjection}
            \begin{tabularx}{\textwidth}{lrrr}
                \toprule
                \textbf{Scenario} & \multicolumn{1}{c}{\textbf{Bad}} & \multicolumn{1}{c}{\textbf{Average}} & \multicolumn{1}{c}{\textbf{Good}} \\
                \midrule
                \textbf{Nr.\ of turns $\bm{N}$} & 3 & 4 & 5 \\
                \textbf{Rejection rate $\bm{R}$} & 0.5 & 0.15 & 0.0 \\
                \midrule
                \textbf{Initial injection} $\bm{C}$ & 572 & 314 & 200 \\
                \textbf{Total uses} $\bm{U}$ & 1000 & 1000 & 1000 \\
                \bottomrule

            \end{tabularx}
        \end{table}
        
\begin{figure}[H]
            \begin{subfigure}{0.48\textwidth}
                \centering
                \includegraphics[width=\textwidth]{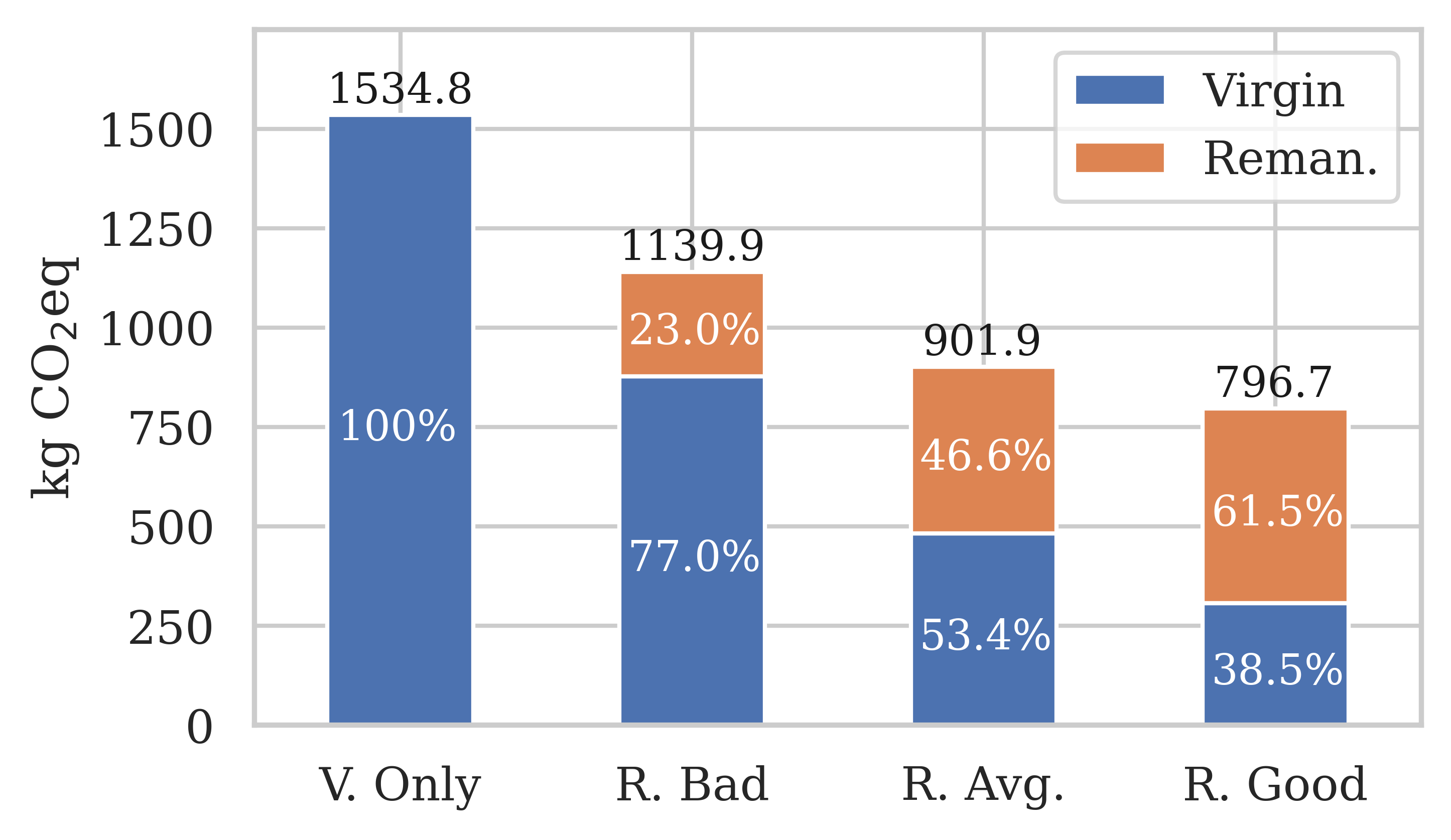}
                \captionsetup{position=bottom,justification=centering}
                \caption{{}}
                \label{fig:buybackEmissionsOverall}
            \end{subfigure}
            \hfill
            \begin{subfigure}{0.48\textwidth}
                \centering
                \includegraphics[width=\textwidth]{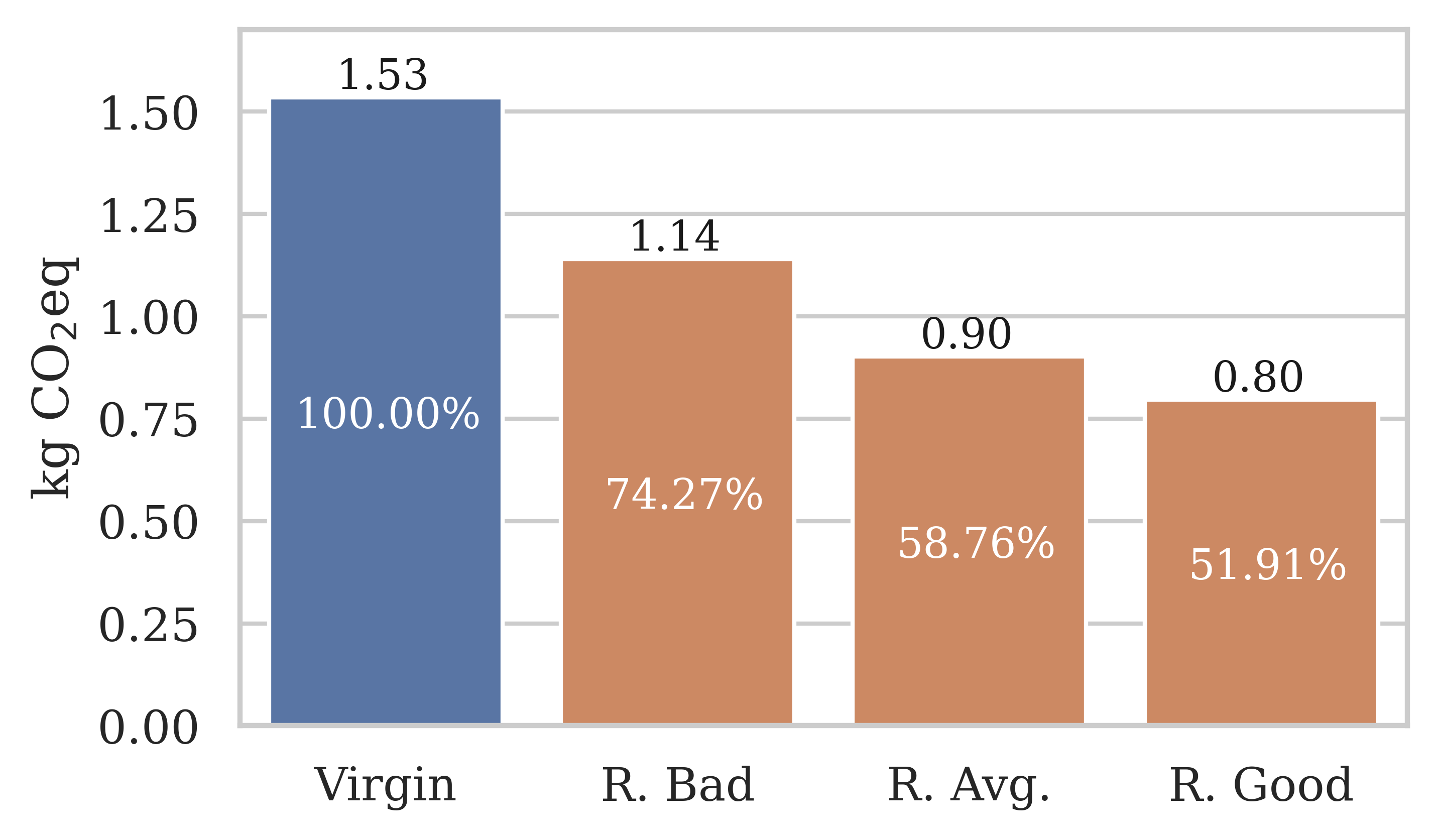}
                \captionsetup{position=bottom,justification=centering}
                \caption{{}}
                \label{fig:buybackEmissionsCatheter}
            \end{subfigure}
            \caption{{\emph{Emission results of a realistic buyback scheme.} A total of 1000 catheter uses $U$ were considered. The proportion of virgin and remanufactured catheter uses are given in Table~\ref{tab:buyBackInjection}. The results show that remanufacturing may reduce long-term emissions by up to 48\%.} (\textbf{a}) Burdened long-term emissions $E^l_{perScheme}$. (\textbf{b}) Burdened long-term emissions $E^l_{perTurn}$.}\label{fig:buybackEmissions}
        \end{figure}

        {Figure~\ref{fig:buybackEmissions}a visualises the total emissions for 1000 catheter uses across scenarios. It immediately stands out that all remanufacturing burdened-emissions $E^b_{perLife}$ were significantly lower than the emissions of 1000 virgin catheters ($E^f_v \cdot 1000$ = 1535 kg CO$_2$eq). As the proportion of uses covered by remanufactured catheters grew, the total emissions reduced from 1140 kg (Bad) to 902 kg (Average) to 797 kg CO$_2$eq (Good).}
        
        {Relative to one remanufactured catheter (Figure~\ref{fig:buybackEmissions}b), the burdened per turn emissions $E^b_{perTurn}$ were reduced to 0.8--1.14 kg CO$_2$eq, compared to a 1.53 kg CO$_2$eq virgin catheter. Overall, our results show that \textbf{remanufacturing may reduce long-term, cumulative emissions by a significant 25--48\%}, depending on the process efficiency and the used catheter~quality.}

\section{Discussion}\label{sec:discussion}
   
\begin{figure}[H]
        \begin{subfigure}{0.522\textwidth}
            \centering
            \includegraphics[width=\textwidth]{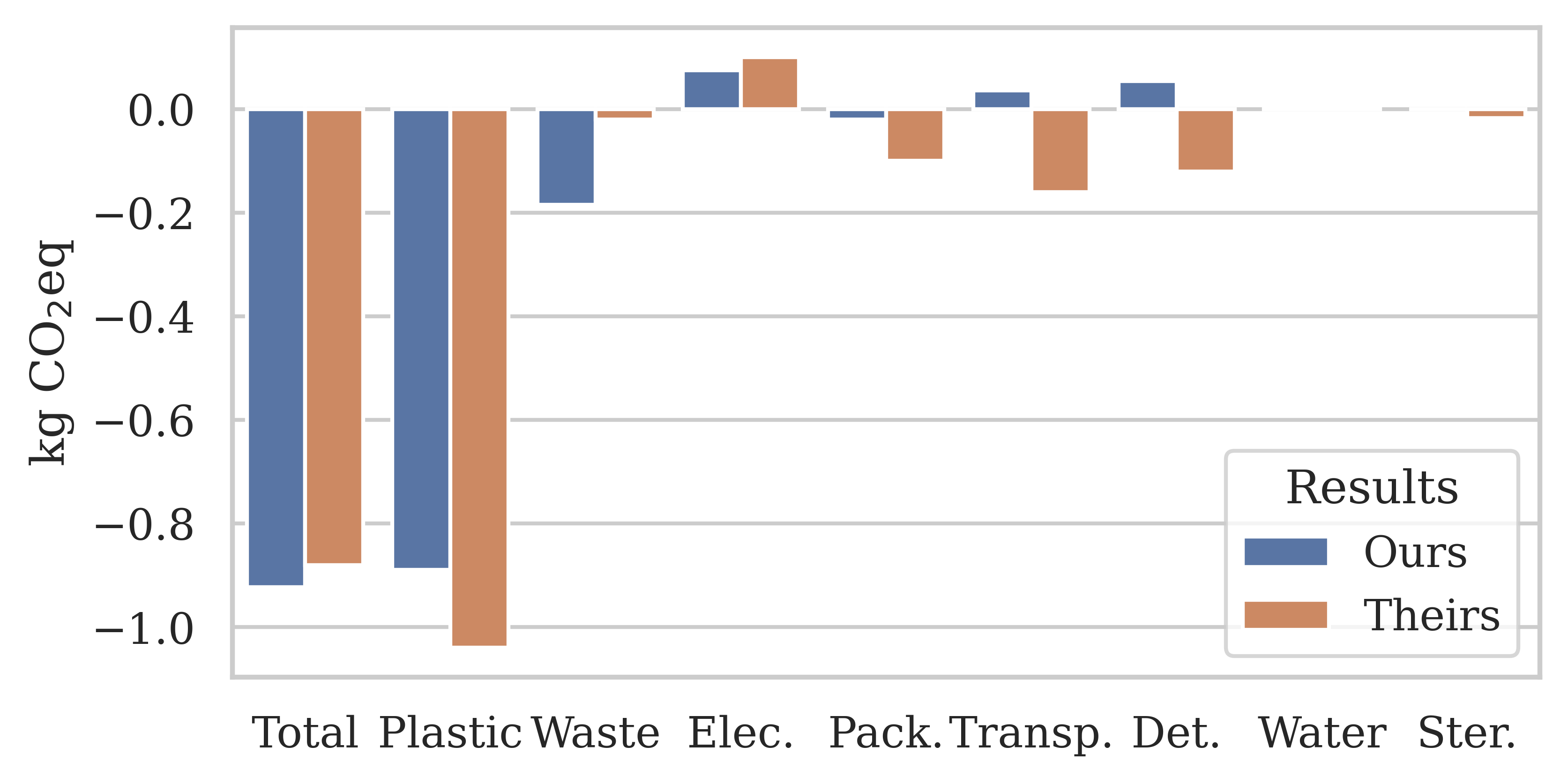}
            \captionsetup{position=bottom,justification=centering}
            \caption{{}}\label{fig:resultDeltas}
        \end{subfigure}
        \hfill
        \begin{subfigure}{0.448\textwidth}
            \centering
            \includegraphics[width=\textwidth]{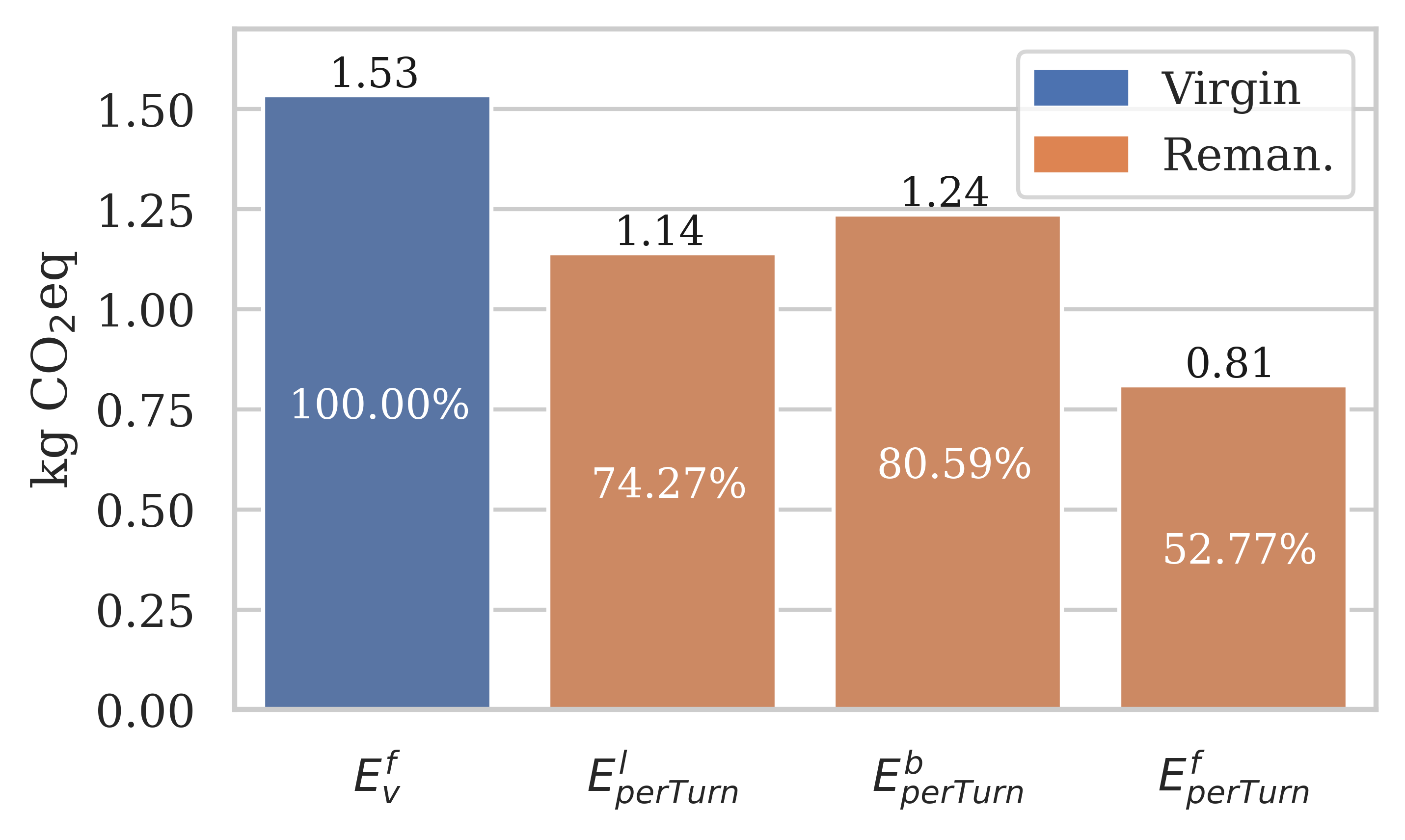}
            \captionsetup{position=bottom,justification=centering}
            \caption{{}}
            \label{fig:UsResultsOverview}
        \end{subfigure}
        \caption{{\emph{A summary of our results.} (\textbf{a}) compares our virgin vs. remanufactured catheter emission results against the case study~\cite{schulte2021combining}. (\textbf{b}) illustrates how much the total remanufacturing emissions may vary depending on the selected emission metric. For a fair comparison, all remanufactured emissions are calculated for the same scenario ($N=3$, $R=50\%$ and $l=USA$, as defined in Section~\ref{sec:sensitivityScenarios}).}}\label{fig:resultSummary}
    \end{figure}

    {Our extensive emission result analysis found that remanufacturing catheters reduces burden-free climate emissions by 60\%. Figure~\ref{fig:resultSummary}a compares our burden-free results $E_v^f = 1.53$ and $E_r^f = 0.61$kg CO$_2$eq (Section~\ref{sec:burdenfreeResults}) against the \emph{Fraunhofer case study}~\cite{schulte2021combining}. The case study reports $E^f_v = 1.75$ and $E^f_r = 0.87$kg CO$_2$eq, representing a 50\% emission reduction through remanufacturing. Even though the total emissions were slightly higher than what our models produced, we consider the results validated as the total and most category emissions follow the same trends. Furthermore, the discrepancies can be explained with the primary data updates we made, such as the lower-impact virgin plastic processing (replacement materials), lower remanufacturing waste emissions (reduced rejection rate) and the increased remanufacturing transport emissions (doubled transport rather than one round trip). Readers are referred to Section~\ref{sec:lci} for details and justifications.}

    {\emph{Emission metrics}.
However, burden-free results do not necessarily accurately reflect the realistic emissions. Figure~\ref{fig:resultSummary}b visualises how the total emissions may change depending on the selected metric. For a fair comparison, all remanufactured emissions are calculated for the same scenario ($N=3$, $R=50\%$ and $l=$USA). Our proposed metric $E_{perTurn}^l$ comes the closest to accurately reflecting the real-world emissions, as it takes the \emph{previous lives} (burdened) and \emph{long-term use} of catheters into account. For the chosen Bad scenario, remanufacturing reduces climate emissions by 47\% burden-free, 19\% burdened, and 26\% long-term, respectively.}
    
    {\emph{Sensitivity analysis.} Of course, these results may improve dramatically when the remanufacturing process is optimised. One of the main insights of our extensive sensitivity analysis (Section~\ref{sec:sensitivityAnalysis}) is the significant impact even relatively minor changes may have on the absolute and cumulative remanufacturing emissions. Higher remanufacturing efficiency (e.g., quality rejection rate, number of turns, location) may significantly reduce climate impact. In our best-case scenario, remanufacturing reduced climate emissions by 71\% burden-free, 57\% burdened, and 52\% long-term, respectively.}

    {\emph{Result uncertainty.}  Because Life Cycle Analysis (LCA) results are calculated from a model of the real-world product, LCA climate emissions should be considered as estimates. The uncertainty of the results may be categorised into three groups: \emph{model}, \emph{scenario}, and \emph{parameter uncertainty}~\cite{bamber2020comparing}. Throughout our study, we have attempted to address these concerns and improve the reliability of our results. Model uncertainty comes from the structure of the emission-calculating models themselves. We have addressed this by choosing highly-regarded software and background datasets openLCA and Ecoinvent v3.0. On the other hand, scenario uncertainty describes uncertainty introduced through LCA methodological choices such as the functional unit and system boundary. To minimise the introduced uncertainty, we have followed industry standards (e.g., ISO~\cite{iso200614040} and NHS guidelines~\cite{penny2012greenhouse,standard2013ghg}). Finally, to address parameter uncertainty (i.e., data uncertainty), we have conducted a qualitative analysis of the data quality and included a comprehensive sensitivity analysis of high-impact life cycle parameters.}

    {\emph{Versatility.}  Note that the systematic evaluation of virgin and remanufactured electrophysiology catheter emissions and the proposed long-term emission metric are versatile and not limited to electrophysiology catheters. In future, a similar approach could be taken to evaluate the life cycle emissions of other high-waste medical devices, as well as for circular economy approaches other than remanufacturing (e.g., recycling).}

\section{Related Work}\label{sec:relatedWork}
      
    {As substantial individual contributors to environmental impacts, healthcare organisations and regulatory bodies worldwide have been investing in circular economy solutions~\cite{england2020delivering,agreement2015paris,eu2020carbon}. Compared to the current reliance on a linear, take-make-dispose model, a circular economy is defined by its principles to redesign, reduce, recover, recycle, and reuse (5R's) resources to significantly extend their life before disposal~\cite{chen2020implementation}. Remanufacturing is a promising technique which resets a used device to a ``substantially equivalent'' state as when it was first manufactured~\cite{mhra2016remanufacture,mhra2021reprocess}. This approach is particularly impactful for single-use medical devices, which contribute significantly to the 590,000 tonnes of healthcare waste produced annually in England alone~\cite{zils2022accelerating} and have similar statistics globally. Regulated remanufacturing processes allow the devices to be legally reused, reducing both the consumption of rare raw resources and minimising the need for emission-heavy waste disposal treatments (e.g., incineration)~\cite{UNGER20161995}.}

    {Apart from the environmental and financial benefits of a more circular single-use device economy, remanufacturing also has the potential to make expensive, multi-use medical devices more available to developing countries~\cite{eze2019accessing}. By prolonging their life, devices discarded due to the fast-paced, technology-driven turnover in high-income countries may be repurposed. Multiple studies have argued that the required initial investment to ramp up remanufacturing capabilities would be offset quickly as the devices are safely reused~\cite{oturu2022remanufacturing,eze2020remanufacturing}.}

    {To make the required sustainable procurement decisions throughout a product value chain, decision makers need access to realistic, quantified climate impacts~\cite{dong2018environmental}. Life Cycle Analysis (LCA) has been widely recognised as an effective tool to support sustainable decision-making by systematically evaluating a product system's potential impacts over its entire life~\cite{liu2022benefit}. Climate impact (also GHG emissions or carbon footprint) measured in kg CO$_2$eq tends to be the most popular emission metric~\cite{smol2017circular,ditac2022carbon}, presumably due to its widespread use across public, industry, and legislative communication around climate change. However, LCAs may be used to give a broad overview of multiple environmental factors instead, which may avoid over-optimising a particular indicator~\cite{bojarski2009incorporating}. The LCA modelling may have different purposes within the scope of sustainability: (i) To compare the impact efficiency of healthcare practices (e.g., healthcare waste management approaches~\cite{soares2013applications}), (ii) to inform product design (e.g., comparing single-use medical devices containing biopolymers vs. plastic equivalents~\cite{unger2017single}, and (iii) to inform procurement decisions (e.g., single-use vs. reusable bronchoscopes~\cite{sorensen2018comparative}). Furthermore, called a Comparative LCA, such analysis may provide a robust and transparent comparison of two or more products~\cite{martin2022lca}.}

    {However, a significant limitation of LCAs is that they rely on a core, approximated model of a real-world product system~\cite{larsson2019consideration}. Therefore, the quality of the environmental impact measurements is subject to the quality of the representative model, which describes all material inputs and outputs that contribute to or stem from the product or service~\cite{wu2019carbon}. This inherent challenge to the modelling process means that LCA models are very time-consuming and expensive to construct, often requiring long periods of extensive data collection throughout a product's life cycle. The consequence is that few studies have been carried out on complex product systems. Popular healthcare candidates tend to be simple with relatively few components, for example, face masks~\cite{van2021life}, surgical scissors~\cite{rizan2022life}, and staplers~\cite{freund2022environmental}, to name a few. In contrast, electrical medical devices are rarely featured (e.g., bronchoscopes~\cite{sorensen2018comparative}). Especially in the healthcare domain, proprietary design and materials in key components may make the data collection and LCA modelling process more complicated than usual.}

    {In cases where product life cycle data is unavailable, assumptions and simplifications have to be made. Because LCAs may be highly sensitive to such choices as well as other modelling decisions (e.g., system boundary, reference unit, allocation strategy)~\cite{djati2018lca,liu2019hybrid,bishop2021environmental}, recent years have seen an attempt to standardise LCA analysis through modelling standards (e.g., ISO regulations~\cite{iso200614040,iso200614044}) and reporting guidelines (e.g., NHS GHG accounting \cite{penny2012greenhouse,standard2013ghg}). Consequently, current studies tend to include a report of the assumptions made~\cite{anshassi2021reviewing,anshassi2021review,kloverpris2018establishing} and a sensitivity analysis of key life cycle parameters~\cite{shimako2018sensitivity,pannier2018comprehensive,patouillard2019prioritizing}. The aim is to give the results and their interpretation a clear context and provide a measure of the uncertainty.}
    
    {Given the uncertainty of LCA estimations and their importance in supporting significant shifts in sustainable medical device procurement, it is not unreasonable to expect studies to be frequently validated to ensure that the results are accurate. However, LCA result validation is rare because comparing absolute environmental impact results across studies is difficult even when common guidelines are followed~\cite{spreafico2022analysis}. This is because the mentioned industry standards propose best practices in broad strokes but leave many details to the study designers, which may drastically influence the results. For example, there is currently no one-size-fits-all solution for incorporating circular life cycles in LCA models, nor a hard-and-fast rule for which life cycle parameters to consider when calculating long vs. short-term emissions. When validation is attempted, it often results in major impact discrepancies (e.g., $-$270--570\% variation in a validation of 13 plastic recycling studies~\cite{creadore2022quantitative}).}

    {The challenges with LCA emission modelling illustrated above also apply to electrophysiology catheters, the medical device examined in this article. As a traditionally single-use medical device with significant rare-resource material components, they are a promising candidate for remanufacturing. To the best of our knowledge, only a handful of studies have assessed their environmental impacts, which we discuss in the following.}

    In a recent study, Leung et al.\ found that circular mapping catheters could be remanufactured to their original functionality and efficiently reused in healthcare centres without complications~\cite{leung2019remanufactured}. According to their calculations, over \textsterling30,000 were saved across only 100 procedures. They estimate that up to \textsterling1.7 m may be saved annually in the UK if half of all circular mapping catheter procedures (5000--10,000) were carried out with remanufactured catheters instead. The paper concludes that catheter remanufacturing is a highly cost-efficient process that may be implemented without compromising patient safety. Additionally, Ditac et al.\ modelled the carbon footprint of atrial fibrillation catheter ablation procedures from catheter production to surgery, covering all processes carried out in the operating room~\cite{ditac2022carbon}. The authors found that catheters contributed almost 40\% of total Greenhouse Gas (GHG) emissions during the procedure, 26\% of emissions came from the anaesthesia and other pharmaceutical drugs, and the remaining 34\% were emitted by other disposable materials during surgery. Since more than a third of ablation GHG emissions are released during material mining and catheter production, the results suggest that catheter remanufacturing could potentially significantly reduce their environmental~impact.

    A survey of 278 physicians from 42 European healthcare centres revealed a high motivation to reduce the environmental impact of EP procedures (62\% of responses)~\cite{boussuge2022current}. However, Boussuge-Roze et al.\ report that only 15\% of catheters are remanufactured and identify multiple barriers that must be addressed before the process can become more widespread. These are mainly at an institutional, processing, and regulatory level and highlight the need for a collaborative approach between healthcare, industry, and legislation. Schulte et al.\ provide an incentive for catheter remanufacturing by quantifying the emission reductions with a comparative LCA analysis of virgin manufactured and remanufactured single-use electrophysiology catheters~\cite{schulte2021combining}. They find that remanufacturing reduces GHG emissions by 51\% per turn when classifying the used catheter at the start of the remanufacturing process as burden-free (i.e., not considering previous lives). They also include a circularity metric to take multiple remanufactured catheter lives into account, which can be critical to interpret LCA results accurately~\cite{andrae50weighting}.

    {Even with widely-accepted standards for LCA analysis in place, LCA methodology for healthcare faces open questions. Significant challenges include (i) how to address data unavailability (e.g., due to proprietary materials and processes), (ii) how to address the consequent uncertainty, and (iii) how to translate the absolute LCA emission results into actionable insights for industry and healthcare practitioners. To address these points in our comprehensive study of electrophysiology catheters, we validate a case study from the literature, conduct an extensive results analysis including a sensitivity analysis of key parameters, and propose a novel metric that calculates realistic emissions by taking long-term use statistics of the device into account.}

\section{Conclusions}
   
    {In this article, we carried out a comprehensive analysis of the climate impact and carbon emissions of virgin manufactured and remanufactured electrophysiology catheters, validating and expanding on a previous case study by the prestigious Fraunhofer Institute with a holistic sensitivity analysis and long-term emission analysis. Our emission framework is based on Life Cycle Analysis (LCA) models.}
    
    {According to our models, remanufacturing catheters achieved \textbf{60\% burden-free per turn emission reductions}
(1.53 to 0.61 kg CO$_2$eq). However, our results have shown that burden-free results do not necessarily reflect real-world emissions. Therefore, we have included a range of emission metrics, including a novel long-term emission metric. When taking a remanufactured catheter's previous lives into account, the emission reduction drops slightly to \textbf{57\% burdened per life reductions} ($-$0.87 kg CO$_2$eq). Including long-term remanufacturing use statistics changes the results to up to \textbf{48\% cumulative reductions} ($-$0.73 kg CO$_2$eq).}
    
    {Our comprehensive results analysis concluded in three primary insights:}
    \begin{enumerate}[label=(\roman*)]
        \item {Remanufacturing is a promising circular economy approach to reduce the climate impact of single-use electrophysiology catheters.}
        \item {Sensitivity analysis has shown that parameters across the product chain may have major impacts on environmental emissions (e.g., up to 66\% reduction with an improved remanufacturing rejection rate). Therefore, our study encourages a collaborative approach to remanufacturing by all actors throughout the value~chain.}
        \item {Furthermore, finally, the distinction between burden-free (no previous lives) and burdened emissions (taking previous lives into account) is necessary to fully understand how emissions may be attributed to catheter turns over its entire life. Additionally, long-term use statistics should be incorporated into the emission metrics for more accurate results.}
    \end{enumerate}
    
    {In future, we would like to expand our LCA models by removing the need for simplified assumptions that we had to make due to data availability. Additionally, since our proposed emissions framework is device-agnostic, it would be beneficial to analyse the emissions of other high-waste medical devices to inform the developing circular economy and net zero goal in healthcare.}

\vspace{6pt} 

%%%%%%%%%%%%%%%%%%%%%%%%%%%%%%%%%%%%%%%%%%
%% optional
%\supplementary{The following supporting information can be downloaded at:  \linksupplementary{s1}, Figure S1: title; Table S1: title; Video S1: title.}


%%%%%%%%%%%%%%%%%%%%%%%%%%%%%%%%%%%%%%%%%%
\authorcontributions{Conceptualization, Y.W. and K.A.N.; methodology, J.A.M., J.S., Y.W. and K.A.N.; software, J.A.M. and J.S.; validation, J.A.M., J.S., Y.W. and K.A.N.; formal analysis, J.A.M. and J.S.; investigation, J.A.M. and J.S.; resources, J.A.M., J.S., Y.W. and K.A.N.; data curation, J.A.M., J.S., Y.W. and K.A.N.; writing---original draft preparation, J.A.M. and J.S.; writing---review and editing, J.A.M., J.S., Y.W. and K.A.N.; visualization, J.A.M. and J.S.; supervision, Y.W. and K.A.N.; project administration, Y.W. and K.A.N.; funding acquisition, Y.W. and K.A.N. All authors have read and agreed to the published version of the manuscript.}

\funding{This research was funded by University of Brighton's 2021 Quality Research grant.}

\institutionalreview{Not applicable.}

\informedconsent{Not applicable.}

\dataavailability{The data presented in this study are available in this article and in the licensed Ecoinvent 3.8 dataset at \url{https://ecoinvent.org/}, accessed on 11 July 2022.}

\acknowledgments{We gratefully thank Dan Vukelich and David Sheon from the Association of Medical Device Reprocessing (AMDR); Lars Thording from Innovative Health; and Agnes Henson, Nicole Fletcher and Rebecca Griffiths from NHS England for their insight and contribution. This research was funded by University of Brighton's 2021 Quality Research grant.}

\conflictsofinterest{The authors declare no conflict of interest. The funders had no role in the design of the study; in the collection, analyses, or interpretation of data; in the writing of the manuscript; or in the decision to publish the~results.}

%%%%%%%%%%%%%%%%%%%%%%%%%%%%%%%%%%%%%%%%%%
%% Optional
% \sampleavailability{Samples of the compounds ... are available from the authors.}

%% Only for journal Encyclopedia
%\entrylink{The Link to this entry published on the encyclopedia platform.}

\abbreviations{Abbreviations}{
The following abbreviations are used in this manuscript:\\

\noindent 
\begin{tabular}{@{}ll}
AMDR & Association of Medical Device Reprocessors (USA). \\
CO$_2$ & Carbon Dioxide. \\
DE & Germany. \\
EF & Environmental Footprint. \\
EP & Electrophyisiology. \\
EU & European Union. \\
FDA & Food and Drug Administration (USA). \\
GHG & Greenhouse Gases. \\
GWP & Global Warming Potential. \\
IH & Innovative Health (USA). \\
ISO & International Organization for Standardization. \\
JIT & Just In Time. \\
LCA & Life Cycle Analysis. \\
LCI & Life Cycle Inventory. \\
LCIA & Life Cycle Impact Analysis. \\
NHS & National Health Service (UK). \\
OEM & Original Equipment Manufacturer. \\
SUD & Single-Use Device. \\
UK & United Kingdom. \\
UN & United Nations. \\
USA & United States of America. \\
\end{tabular}
}

%%%%%%%%%%%%%%%%%%%%%%%%%%%%%%%%%%%%%%%%%%
%% Optional
% \appendixtitles{no} % Leave argument "no" if all appendix headings stay EMPTY (then no dot is printed after "Appendix A"). If the appendix sections contain a heading then change the argument to "yes".
% \appendixstart
% \appendix
% \section[\appendixname~\thesection]{}\label{app:lci}
% \input{app_lci}
% \subsection[\appendixname~\thesubsection]{}

% The appendix is an optional section that can contain details and data supplemental to the main text---for example, explanations of experimental details that would disrupt the flow of the main text but nonetheless remain crucial to understanding and reproducing the research shown; figures of replicates for experiments of which representative data are shown in the main text can be added here if brief, or as Supplementary Data. Mathematical proofs of results not central to the paper can be added as an appendix.

% \begin{table}[H] 
% \caption{This is a table caption.\label{tab5}}
% \newcolumntype{C}{>{\centering\arraybackslash}X}
% \begin{tabularx}{\textwidth}{CCC}
% \toprule
% \textbf{Title 1}	& \textbf{Title 2}	& \textbf{Title 3}\\
% \midrule
% Entry 1		& Data			& Data\\
% Entry 2		& Data			& Data\\
% \bottomrule
% \end{tabularx}
% \end{table}

% \section[\appendixname~\thesection]{}
% All appendix sections must be cited in the main text. In the appendices, Figures, Tables, etc. should be labeled, starting with ``A''---e.g., Figure A1, Figure A2, etc.

\FloatBarrier{}
%%%%%%%%%%%%%%%%%%%%%%%%%%%%%%%%%%%%%%%%%%
\begin{adjustwidth}{-\extralength}{0cm}

\reftitle{References}

%%%%%%%%%%%%%%%%%%%%%%%%%%%%%%%%%%%%%%%%%%
\PublishersNote{}

\end{adjustwidth}
\end{document}